\documentclass[journal]{IEEEtran}
\pdfoutput=1
\usepackage{multirow}
\usepackage{booktabs}
\usepackage{amsfonts}
\usepackage{amsmath,cases}
\usepackage{amssymb}
\usepackage{graphicx}
\usepackage{cite}
\usepackage{epsfig}
\usepackage{url}
\usepackage{algorithm}
\usepackage{algorithmicx, algpseudocode}
\usepackage{epstopdf}
\usepackage{color}
\usepackage[tight]{subfigure}
\newcommand{\clA}{{\cal A}}

\newcommand{\ds}{\displaystyle}

\newcommand{\la}{\langle}
\newcommand{\ra}{\rangle}

\newcommand{\bX}{\boldsymbol{X}}
\newcommand{\bx}{\boldsymbol{x}}

\newcommand{\clL}{{\cal L}}
\newcommand{\clN}{{\cal N}}

\newcommand{\clK}{{\cal K}}

\newcommand{\clF}{{\cal F}}

\newcommand{\bv}{\boldsymbol{v}}

\newcommand{\bp}{\boldsymbol{p}}
\newcommand{\pik}{p^{i,(\kappa)}}

\newcommand{\xik}{x^{i,(\kappa)}}

\newcommand{\clB}{{\cal B}}

\newcommand{\by}{\boldsymbol{y}}

\newcommand{\clH}{{\cal H}}

\newcommand{\bY}{\boldsymbol{Y}}

\newcommand{\tbx}{\tilde{\boldsymbol{x}}}

\newcommand{\bV}{\boldsymbol{V}}

\newcommand{\Row}{{\sf Row}}

\newcommand{\tpi}{\tilde{\pi}}

\newcommand{\xk}{x^{(\kappa)}}

\newcommand{\bbY}{\bar{Y}}

\newcommand{\Bk}{B^{(\kappa)}}
\newcommand{\Ck}{C^{(\kappa)}}
\newcommand{\ak}{a^{(\kappa)}}
\newcommand{\pk}{p^{(\kappa)}}
\newcommand{\bk}{b^{(\kappa)}}
\newcommand{\ck}{c^{(\kappa)}}
\newcommand{\rk}{r^{(\kappa)}}
\newcommand{\mk}{m^{(\kappa)}}
\newcommand{\pko}{p^{(\kappa+1)}}
\newcommand{\xko}{x^{(\kappa+1)}}
\newcommand{\muk}{\mu^{(\kappa)}}
\newcommand{\rhok}{\rho^{(\kappa)}}
\newcommand{\clG}{\mathcal{G}}
\allowdisplaybreaks
\begin{document}
\title{A New Class of Structured Beamforming for Content-Centric Fog Radio Access Networks}
\author{W. Zhu$^{1,2}$, H. D. Tuan$^2$, E. Dutkiewicz$^2$,  Y. Fang$^1$, and L. Hanzo$^3$
\thanks{This work was supported in part by the Australian Research Council's Discovery Projects under Grant DP190102501, in part by the Engineering and Physical Sciences Research Council projects EP/P034284/1 and EP/P003990/1 (COALESCE), and in part by the European Research Council's Advanced Fellow Grant QuantCom (Grant No. 789028) }
\thanks{$^1$School of Communication and Information Engineering, Shanghai University, Shanghai 200444, China;
Email: wenbozhu@shu.edu.cn, yfang@staff.shu.edu.cn}
\thanks{$^2$School of Electrical and Data Engineering, University of Technology Sydney, Broadway, NSW 2007, Australia; Email: wenbo.zhu@student.uts.edu.au, tuan.hoang@uts.edu.au, eryk.dutkiewicz@uts.edu.au}
\thanks{$^3$School of Electronics and Computer Science, University of Southampton, Southampton, SO17 1BJ, U.K; Email: lh@ecs.soton.ac.uk }
}
\date{}
\maketitle

\begin{abstract}
A multi-user fog radio access network (F-RAN) is designed for supporting content-centric services. The requested contents are partitioned into sub-contents, which are then `beamformed' by the remote radio heads (RRHs) for transmission to the users. Since a large number of beamformers must be designed, this poses a computational challenge. We tackle this challenge by proposing a new class of regularized zero forcing beamforming (RZFB) for directly mitigating the inter-content interferences, while the `intra-content interference' is mitigated by successive interference cancellation at the user end. Thus each beamformer is decided by a single real variable (for proper Gaussian signaling) or by a pair of complex variables (for improper Gaussian signaling). Hence the total number of decision variables is substantially reduced to facilitate
tractable computation. To address the problem of energy efficiency optimization subject to multiple constraints, such as individual user-rate requirement and the fronthauling constraint of the links between the RRHs and the centralized baseband signal processing unit, as well as the total transmit power budget, we develop low-complexity path-following algorithms. Finally, we actualize their performance by simulations.
\end{abstract}

\section{INTRODUCTION} \label{sec:Intro}
Fog radio access networks (F-RANs) \cite{Peetal15,CZ16,Peetal16,ku20175g,Bitetal17}, which place processing units at the network edge for reducing the backhaul's latency and traffic load,
have emerged for meeting
the radical requirements of mission-critical cloud radio access networks (C-RANs), as exemplified by augmented reality/virtual reality (AR/VR), device-to-device (D2D) communications,  smart living and smart cities, etc. (see e.g.
\cite{Zhuetal15,Mouetal18,DP19,KSS20} and references therein). On one hand,
the uniform quality of service (QoS) provision for cell-edge and cell-center users is satisfied by exploiting the strategic spatial distribution of the remote radio heads (RRHs) over the network
\cite{Pasetal18,Azietal18,Heetal19,YPYH20}. On the other hand, employing RRHs for proactively  caching popular contents enables flawless low-latency service provision relying on a high network throughput at a high energy efficiency, because only a small fraction of the requested video clips must be fetched through the limited-capacity fronthaul links \cite{wang2014cache,LY16,Yanetal16,FZ16}.

Beamforming aided RRHs have been considered in \cite{tao2016content,PSS16,Taoetal19,Huyetal20,Wanetal20} and in the references therein. However, this design problem is high-dimensional, as it involves many beamforming vectors calculated for covering specific segments of contents.
The resultant problem is nonconvex because the throughput function is neither convex nor concave, hence making the objection function constructed for maximizing the sub-contents' throughput nonconcave, while the associated fronthaul constraints are nonconvex. The challenge of dimension is even further escalated in \cite{tao2016content,PSS16}, which treat beamforming vectors of dimension $n_t$ ($n_t$ is the number of RRH antennas), hence ultimately arriving at rank-one matrices of dimension $n_t\times n_t$. The resultant multiple rank-one constraints are then dropped for the sake of facilitating so-called difference of two convex functions based iterations \cite{KTN12}. Despite dropping these rank-one constraints, the problem still remains computationally complex, since it relies on logarithmic determinant function optimization, which has unknown computational complexity. In \cite{tao2016content}, no complexity analysis was provided, while the numerical examples in \cite{PSS16} are limited to the simplest and lowest-dimensional case of three single-antenna RRHs ($n_t=1$) serving three users. The path-following algorithms proposed in \cite{Huyetal20} handled more practical scenarios of five four-antenna RRHs serving five users, which became possible as a benefit of the structural exploitation of approximate zero-forcing beamforming. In contrast to the iterations used in \cite{tao2016content,PSS16}, the path-following algorithms of \cite{Huyetal20} invoke a convex fractional solver of polynomial complexity at each iteration. However, it still remained an open challenge to reduce the dimension of the approximate zero forcing beamformers, while maintaining the throughput.

Against the above background, this paper offers the following new contributions to the design of RRHs beamformers.
\begin{itemize}
\item We propose a new class of regularized zero forcing beamformers (RZFB) operating in the presence of both `intra-content' and `inter-content' interferences, which allows us to recast the beamforming design into an optimization problem of moderate dimension.
\item We conceive a new convex quadratic approximation for the fronthaul constraints, which allows us to design new path-following
algorithms for determining the beamforming vectors by relying on a low-complexity convex quadratic solver at each iteration for generating an improved feasible point.
\end{itemize}
The paper is organized as follows. Section II is devoted to the modeling of the constrained-backhaul FRAN. Sections III and IV constitute the main technical contribution of the paper, which are respectively devoted to RZFB-based proper Gaussian signaling (PGS) and improper Gaussian signaling (IGS). Our simulations are discussed in Section V, while our conclusions offered in Section VI. The Appendix provides fundamental inequalities, which support the derivation of the technical results.

\emph{Notation and most frequently used mathematics. \;} Only the optimization variables are boldfaced to emphasize their
appearance in nonlinear functions; $|{\cal A}|$ is the cardinality of the set ${\cal A}$;
$[X]^2\triangleq XX^H$ and $\la X\ra$ is the trace of the matrix $X$;
${\cal CN}(0,a)$ is the set of circular Gaussian random variables with zero mean and variance $a$
(each $s\in {\cal CN}(0,a)$ is also called proper because $\mathbb{E}(s^2)=0$), while ${\cal C}(0,a)$ is the set of non-circular Gaussian random variables with zero means and variance $a$ ($\mathbb{E}(s^2)\neq 0$ for $s\in {\cal C}(0,a)$ so is also called improper);
 $\mbox{diag}[x]$
for vector $x$ is a diagonal matrix with the entries of $x$ as its diagonal entries; ${\cal I}(x,y)$ is the mutual
information between the random variable $x$ and random variable $y$; The dot product of
the matrices $X$ and $Y$ of appropriate size is defined as $\la X,Y\ra\triangleq \la X^HY\ra$; $X\succ 0$ ($X\succeq 0$, resp.) means $X$ is a Hermitian symmetric ($X=X^H$) and positive definite (positive semi-definite) matrix. Accordingly,
$X\succeq Y$ merely means $X-Y\succeq 0$. One of the most fundamental properties of positive definite matrices is
$X\succeq Y\succ 0\Leftrightarrow Y^{-1}\succeq X^{-1}\succ 0$ \cite{HJ85}.
$||.||$ is the Frobenius  norm, i.e. $||X||=\sqrt{\la [X]^2\ra}$, which also implies $\la A,[X]^2\ra=||A^{1/2}X||^2$ whenever $A\succeq 0$. Therefore, the function $f(X)=\la A,[X]^2\ra +\la B,X\ra$ for $A\succeq 0$ is termed as convex quadratic, while $-f(X)$ is concave quadratic.
\section{FRAN modeling and signaling}
\begin{figure}[!t]
    \centering
	\includegraphics[width=7.4cm]{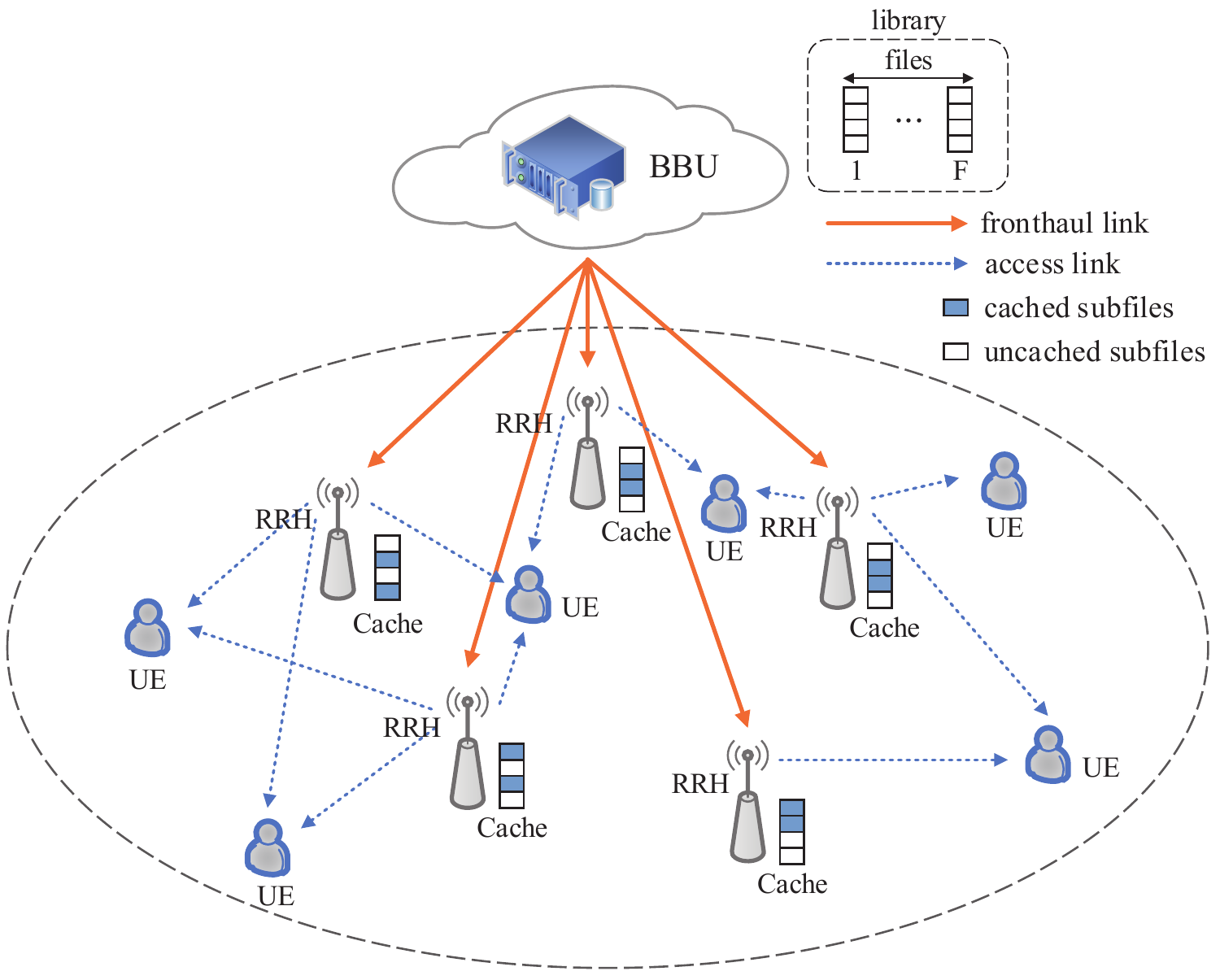}
	\caption{F-RAN}
	\label{fig:model}
\end{figure}

As illustrated by Fig. \ref{fig:model}, we consider a typical F-RAN consisting of a centralized baseband signal processing unit (BBU) and $N_R$ uniformly distributed RRHs indexed by $i\in \clN_R\triangleq \{1,\dots, N_R\}$, which provides content delivery for $N_u$ randomly localized users (UEs) indexed by $k\in \clN_u\triangleq \{1,\dots, N_u\}$. Each RRH is equipped by  an $n_t$-antenna array, while each user equipment (UE) has a single antenna.
The RRHs are connected to the BBU through fronthaul links, each of which is of capacity $C$.

\subsection{Edge caching}
The file library consists of $F$ files  $f\in\clF\triangleq \{1,\dots, F\}$ labelled in order of their popularity, which is  distributed according to Zipf's distribution obeying \cite{breslau1999web} $P(f)=f^{-\gamma_z}/\sum_{j\in\clF}j^{-\gamma_z}$ with
the popularity exponent being $\gamma_z>0$. A larger $\gamma_z$ results in a reduced number of extremely popular contents.
Under the uncoded strategy, each file $f$ is split into $L$ subfiles $(f,\ell)$, $\ell\in\clL\triangleq \{1,\dots, L\}$. Each RRH can store a fraction $\mu=N_c/F_c$ of each
file in the pre-fetching phase \cite{STS17}, where $F_c$ is the library capacity. We define the binary-indicator function associated with  $c^i_{f,\ell}$ such that
$c^i_{f,\ell}=1$ if and only if $(f,\ell)$ is cached by RRH $i$.\footnote{$c^i_{f,\ell}$ is thus pre-determined by the caching strategy RRH used}
Then we have:
\[
\sum_{f=1}^F\sum_{\ell=1}^Lc^i_{f,\ell}/L\leq \mu F,\ i\in\clN_R.
\]

\subsection{Content transmission and reception}
Let us denote the set of requested files by $\clF_{req}$:
\[
\clF_{req}\triangleq \{f_\nu\in\clF :\ \nu=1,\dots, N_{req}\},
\]
where $N_{req}$  ($1\leq N_{req}\leq N_u$) is the number of requested files.
For each $\nu$, we define $\clK_\nu$ as the set of UEs requesting file $f_\nu$:
\begin{equation}\label{knu}
\clK_\nu\triangleq \{1_{\nu},\dots, K^{(\nu)}\},
\end{equation}
where it is plausible that $K^{(\nu)}$ is equal to the cardinality $|\clK_\nu|$ of
$\clK_\nu$.

In what follows, for convenience of presentation, we use the notation $c^i_{\nu,\ell}$ to represent the above
$c^i_{f_\nu,\ell}$ and $\clN_{req}\triangleq\{1,\dots, N_{req}\}$.

Furthermore, the subfile $(f_\nu,\ell)$ requested by UE $k_\nu$ is transferred by the fronthaul links to those $N_F$ RRHs, which have the highest
channel gains from them to UE $k_\nu$ among the RRHs, but do not have $(f_\nu,\ell)$ stored in their cache.
Given $(f_\nu,\ell)$, let us define $d^i_{\nu,\ell}$ as:
\begin{equation}\label{del1}
d^i_{\nu,\ell}\!=\!\begin{cases}\!\begin{array}{ll} \!1&\!\!\mbox{\!if subfile}\ (f_\nu,\ell)\ \mbox{is transferred to RRH}\ i, \cr
\!0&\!\!\mbox{\!otherwise}.
\end{array}
\end{cases}
\end{equation}
For $i=1, \dots, N_R$, we denote by $\clN_i$ the specific set of subfiles that
are either in the cache of RRH $i$ or are received by RRH $i$ from the BBU:
\begin{equation}\label{storage}
\clN_i\triangleq \{(\nu,\ell) : \nu\in \clN_{req},\ \mbox{and}\ c^i_{\nu,\ell}=1\ \mbox{or}\ d^i_{\nu,\ell}=1\}.
\end{equation}
Let $s_{\nu,\ell}\in {\cal CN}(0,1)$ be the proper Gaussian information source that encodes the subfile $(f_\nu,\ell)$.
Each $s_{\nu,\ell}$ for $(\nu,\ell)\in\clN_i$ is `beamformed' into the signal $\lambda^i_{\nu,\ell}(s_{\nu,\ell})$ for transmission.

\begin{table*}[!t]
   \centering
   \caption{Nomenclature}
   \begin{tabular}{|c|r|}
    \hline
   Notation & \multicolumn{1}{c|}{Description}    \\
   \hline\hline
   $f_{\nu}$&a requested file \\ \hline
   $(f_{\nu},\ell)$& the $\ell$-th subfile of file $f_{\nu}$ \\ \hline
    $s_{\nu,\ell}$& the information source encoding $(f_{\nu},\ell)$\\ \hline
    ${\cal K}_{\nu}$& the set of users requesting file $f_{\nu}$ (defined by (\ref{knu}))\\ \hline
    $|{\cal K}_{\nu}|$& the number of users requesting file $f_{\nu}$\\ \hline
    $k_{\nu}$& user $k$ requests file $f_{\nu}$\\ \hline
   $G^i_{k_\nu,\ell}$& transmit beamformer for $s_{\nu,\ell}$ at RRH $i$ by the user $k_\nu$'s request
   (defined by  (\ref{rec1}))\\ \hline
  $\bp^{i}_{k_\nu,\ell}$& optimization variable for allocating the power to $G^i_{k_\nu,\ell}$ in PGS (\ref{prec2a})\\ \hline
  $(\bp^{i,1}_{k_\nu,\ell}, \bp^{i,1}_{k_\nu,\ell})$& optimization variable for allocating the power to $G^i_{k_\nu,\ell}$ in IGS (\ref{rec2a})\\ \hline
  $\upsilon_i$& quantization noise in BBU transmission to RRH $i$\\ \hline
  $\bx^i$ & the introduced variable to control $\upsilon$ (defined by (\ref{bx}))\\ \hline
\end{tabular}
\label{tab:Notation}
\end{table*}

Let $\Xi^i\triangleq \clN_i\setminus \{(\nu,\ell) : c^i_{\nu,\ell}=1\}$.
Under the soft-transfer fronthauling (STF) regime of \cite{Paetal13}, the BBU transfers the following quantization-contaminated version of the `beamformed' subfiles to RRH $i$:
\begin{eqnarray}
\xi_i=\sum_{(\nu,\ell)\in\Xi^i}\lambda^i_{\nu,\ell}(s_{\nu,\ell})+\upsilon_i,\label{ist3a}
\end{eqnarray}
with the independent quantization noise given by
$\upsilon_i\in{\cal C}(0,\Delta^2(\tbx^i)+\epsilon I_{n_t})$ and
\begin{equation}\label{ist4}
\Delta(\bx^i)\triangleq \mbox{diag}[\tbx^i]\in\mathbb{R}^{n_t\times n_t}, \tbx^i\triangleq (\tbx^i_1,\dots, \tbx^i_{n_t})^T\in\mathbb{R}^{n_t}_+.
\end{equation}
The fronthaul-rate constraint is constituted by the following reliably recovered rate-constraint of SFT \cite{Paetal13}
\begin{equation}\label{mut}
{\cal I}\left(\xi_i,\sum_{(\nu,\ell)\in\Xi^i}\lambda^i_{\nu,\ell}(s_{\nu,\ell})\right)\leq C, i\in\clN_R.
\end{equation}
It is plausible that a finer quantization results in a reduced error covariance $\Delta(\bx^i)$, hence
leading to a higher throughput, but the soft-transfer of the quantized signals is limited by the capacity of STF according to (\ref{mut}). As analysed in detail in \cite{Huyetal20},
SFT is much more efficient than hard-transfer fronthauling (HFT) in the face of limited fronthauling capacity.

Furthermore, the signal transmitted by RRH $i$ is
\begin{equation}\label{chi}
\chi_i\triangleq \sum_{(\nu',\ell)\in\clN_i}\lambda^{i}_{\nu',\ell}(s_{\nu',\ell})
+\upsilon_i,
\end{equation}
which contains the quantization noise $\upsilon_i$. This noise cannot be completely eliminated due to the limited fronthaul-rate constraint (\ref{mut}).  The signal received by UE $k_{\nu}$ is
\begin{align}
y_{k_\nu}&=\ds \sum_{i=1}^{N_R}h_{k_\nu,i}^H\chi_i+ z_{k_\nu}\nonumber\\
&=\ds \sum_{i=1}^{N_R}h_{k_\nu,i}^H\left(\sum_{(\nu',\ell)\in\clN_i}\lambda^{i}_{\nu',\ell}(s_{\nu',\ell})  \right)+ z_{k_\nu},
\label{rec}
\end{align}
where $h_{k_\nu,i}\in\mathbb{C}^{n_t}$ is the channel spanning from RRH $i$ to UE $k_\nu$, and
$z_{k_\nu}\in\mathbb{C}$ is the background noise of covariance $\sigma$.

\section{Regularized zero-forcing beamforming for proper Gaussian signaling}
To assist the reader,
Table \ref{tab:Notation} provides a summary of the key notations at a glance.

\subsection{PGS Problem statement}
It follows from (\ref{rec}) that
the interference of $\lambda^i_{\nu,\ell}(s_{\nu,\ell})$ to UE $k_{\nu'}$  ($\nu'\neq \nu$) is given by
$h^H_{k_{\nu'},i}\lambda^i_{\nu,\ell}(s_{\nu,\ell})$. Thus, for $n_{\nu}\triangleq \sum_{\nu'\in\clN_{req}\setminus\{\nu\}}|\clK_{\nu'}|$,
we define the matrix of interfering channels by
\begin{equation}\label{uns1}
\Phi^i_{k_\nu,\ell}\triangleq \Row (h^H_{k_{\nu'},i})_{\nu'\in\clN_{req}\setminus\{\nu\}, k=1,\dots, |\clK_{\nu'}|}
\in\mathbb{C}^{n_{\nu}\times n_t},
\end{equation}
and then the matrix of signal plus interfering channels as
\begin{equation}\label{uns1a}
\tilde{\Phi}^i_{k_\nu,\ell}\triangleq \begin{bmatrix}h^H_{k_{\nu},i}\cr
\Phi^i_{k_\nu,\ell}
\end{bmatrix}\in\mathbb{C}^{(n_{\nu}+1)\times n_t}.
\end{equation}
Here the operation $\Row$ in (\ref{uns1}) arranges $n_{\nu}$ row-vectors
$h^H_{k_{\nu'},i}\in\mathbb{C}^{1\times N_t}$,  $\nu'\in\clN_{req}\setminus\{\nu\}$, $k=1,\dots, |\clK_{\nu'}|$ in  a matrix of $n_{\nu}$ rows.

For $(\nu,\ell)\in\clN_i$ and
\begin{equation}\label{rec1}
G^i_{k_\nu,\ell}\triangleq \left((\tilde{\Phi}^i_{k_\nu,\ell})^H\tilde{\Phi}^i_{k_\nu,\ell}+\alpha I_{n_t}\right)^{-1}h_{k_{\nu},i}, k=1,\dots, |\clK_\nu|,
\end{equation}
associated with\footnote{$P$ is the power budget}
\begin{equation}\label{alpha}
\alpha=\begin{cases}\begin{array}{ll}\!0&\mbox{if}\ \mbox{rank}(\tilde{\Phi}^i_{k_\nu,\ell})=n_{\nu}+1\cr
\!N_R\sigma/P &\mbox{otherwise},
\end{array}
\end{cases}
\end{equation}
we propose the following class of RZFB
\begin{align}
\lambda^i_{\nu,\ell}(s_{\nu,\ell})&=\sum_{k=1}^{|\clK_{\nu}|}\bp^{i}_{k_\nu,\ell}G^i_{k_\nu,\ell}s_{\nu,\ell} \label{prec2a}\\
&=\left(\sum_{k=1}^{|\clK_{\nu}|}\bp^{i}_{k_\nu,\ell}G^i_{k_\nu,\ell}\right)s_{\nu,\ell},
\ \bp^i_{k_\nu,\ell}\in\mathbb{R}.\label{prec2}
\end{align}
Let $e_1\in\mathbb{R}^{n_\nu+1}$ having all-zero entries except for the first entry, which is $1$. Then we have
\begin{align}
\tilde{\Phi}^i_{k_\nu,\ell}G^i_{k_\nu,\ell}&=\tilde{\Phi}^i_{k_\nu,\ell}\!
\left(\!(\tilde{\Phi}^i_{k_\nu,\ell})^H\tilde{\Phi}^i_{k_\nu,\ell}+\alpha I_{n_t}\!\right)^{\!-1}\!(\tilde{\Phi}^i_{k_\nu,\ell})^He_1\nonumber\\
&=\tilde{\Phi}^i_{k_\nu,\ell}(\tilde{\Phi}^i_{k_\nu,\ell})^H\!\left(\!\tilde{\Phi}^i_{k_\nu,\ell}(\tilde{\Phi}^i_{k_\nu,\ell})^H\!+\!
\alpha I_{n_\nu+1}\!\right)^{\!-1}\!e_1\nonumber\\
&\approx I_{n_\nu+1}e_1\label{rec1a}\\
&=e_1,\label{rec1b}
\end{align}
implying that the power of $\lambda^i_{\nu,\ell}(s_{\nu,\ell})$  can be still amplified,
while its interference inflicted upon UE $k_{\nu'}$  ($\nu'\neq \nu$) is approximately forced to zero. When the rank of
$\tilde{\Phi}^i_{k_\nu,\ell}$ is $(n_{\nu}+1)$, which only occurs for $n_t> n_{\nu}+1$,
 the perfect zero-interference condition of $\lambda^i_{\nu,\ell}(s_{\nu,\ell})$ is achieved.

It should be noted that (\ref{prec2}) represents a brand new class of RZFB specifically tailored for mitigating the inter-content interference only and as such it has quite a different structure compared to the traditional RZFB (see e.g. \cite{Ngetal19}), which aims for suppressing the multi-user interference.

Under the RZFB of (\ref{prec2a}), the signal $\xi_i$ defined in (\ref{ist3a}) and transferred from the BBU to RRH $i$ through the limited-rate backhaul link is reformulated as:
\begin{equation}
\xi=\sum_{(\nu,\ell)\in\Xi^i}\left(\sum_{k=1}^{|\clK_\nu|}\bp^{i}_{k_\nu,\ell}G^i_{k_\nu,\ell}\right)s_{\nu,\ell}
+\upsilon_i.\label{pist3}
\end{equation}
Then, the mutual information (MI) in the right-hand side (RHS) of (\ref{mut}) may be expressed as:
\begin{align}
&{\cal I}\left(\xi_i,\sum_{(\nu,\ell)\in\Xi^i}\left(\sum_{k=1}^{|\clK_\nu|}\bp^{i}_{k_\nu,\ell}G^i_{k_\nu,\ell}\right)s_{\nu,\ell}\right)\nonumber\\
&\quad=\ln\left|I_{n_t}+L_i(\bp)\left(\left[\Delta(\tbx^i)\right]^2+\epsilon I_{n_t}\right)^{-1}\right|\nonumber\\
&\quad=\ln\left|I_{n_t}+\epsilon^{-1}L_i(\bp)\left(\left[\Delta(\bx^i)\right]^2+I_{n_t}\right)^{-1}\right|\nonumber\\
&\quad\triangleq m_i(\bp,\bx^i),\label{pist7}
\end{align}
for $\bp\triangleq \{\bp^i_{k_\nu,\ell}): (\nu,\ell)\in\clN_i, k_\nu=1,\dots, |\clK_\nu|, i=1,\dots, N_R\}$,
and
\[
L_i(\bp)\triangleq \sum_{(\nu,\ell)\in\Xi^i}\left[\sum_{k=1}^{|\clK_\nu|}\bp^{i}_{k_\nu,\ell}G^i_{k_\nu,\ell}\right]^2\in\mathbb{C}^{n_t\times n_t},
\]
and
\begin{equation}\label{bx}
\bx^i=\tbx_i/\sqrt{\epsilon}.
\end{equation}
Then the fronthaul-rate constraint (\ref{mut}) becomes:
\begin{equation}\label{pist8}
m_i(\bp,\bx^i)\leq \log_2(e)C, i\in\clN_R.
\end{equation}
Observe that the unit in (\ref{pist7}) is nat, while the unit of $C$ is bit, so the factor $\log_2(e)$ in
(\ref{pist8}) converts its right-hand side (RHS) to nats for consistency with its left-hand side (LHS).

The transmit signal at RRH $i$ is specialized to
\begin{equation}\label{pist8a}
\chi_i=\sum_{(\nu,\ell)\in \clN_i}\left(\sum_{k=1}^{|\clK_\nu|}\bp^{i}_{k_\nu,\ell}G^i_{k_\nu,\ell}\right)s_{\nu,\ell}+\upsilon_i.
\end{equation}
For each $\nu\in\clN_{req}$ let $\iota_{\nu}$ be the set of RRHs that do not have the entire file $\nu$ in
their cache, i.e.
\begin{equation}\label{ist8a}
\iota_{\nu}\triangleq \{i\in\clN_{R}\ :\ c^i_{\nu,\ell}\not\equiv 1\}.
\end{equation}
The power consumption of transmitting the signal carrying $f_{\nu}$ from RRH $i$ is expressed as:
\begin{eqnarray}
\sum_{\ell: (\nu,\ell)\in\clN_i}\!||\sum_{k=1}^{|\clK_{\nu}|}\bp^{i}_{k_\nu,\ell}G^i_{k_\nu,\ell}||^2\!+\!
\sum_{i\in\iota_{\nu}}\sum_{j=1}^{n_t}\left[(\tbx^i_j)^2+\epsilon\right]= \nonumber\\
\sum_{\ell: (\nu,\ell)\in\clN_i}\!||\sum_{k=1}^{|\clK_{\nu}|}\bp^{i}_{k_\nu,\ell}G^i_{k_\nu,\ell}||^2\!+\!
\epsilon\sum_{i\in\iota_{\nu}}\sum_{j=1}^{n_t}\left[(\bx^i_j)^2+1\right]\triangleq \nonumber\\
\pi^i_{\nu}(\bp,\bx),\label{pist8b}
\end{eqnarray}
for $\bx\triangleq \{\bx^i, i=1,\dots, N_R\}$.  The total power dissipated by delivering $f_{\nu}$ is given by
the following convex quadratic function
\begin{equation}\label{pist9}
\pi_{\nu}(\bp,\bx)\triangleq \sum_{i=1}^{N_R}\pi^i_{\nu}(\bp,\bx) +P_{non},
\end{equation}
where $P_{non}$ is the total `non-transmission' power dissipation at the RRHs formulated as
$N_R n_t P_a$, where the antenna circuit power is $P_a$.

Let us now rewrite equation (\ref{rec}) of the signal received by user $k_\nu$ as
\begin{align}
y_{k_\nu}&=\sum_{\ell=1}^L\sum_{i: (\nu,\ell)\in\clN_i}h_{k_\nu,i}^H\lambda^i_{\nu,\ell}(s_{\nu,\ell})\nonumber\\
&\quad+\sum_{\nu'\in\clN_{req}\setminus\{\nu\}}\sum_{\ell=1}^L\sum_{i:(\nu',\ell)\in\clN_i}\lambda^i_{\nu',\ell}(s_{\nu',\ell})\nonumber\\
&\quad+\sum_{i=1}^{N_R}h_{k_\nu,i}^H\upsilon_i+ z_{k_\nu}\nonumber\\
&=\ds\sum_{\ell=1}^La_{k_\nu,\ell}(\bp)s_{\nu,\ell}
+\sum_{\nu'\in\clN_{req}\setminus\{\nu\}}\sum_{\ell=1}^Lb_{k_\nu,\ell,\nu'}(\bp)s_{\nu',\ell}\nonumber\\
&\quad+\ds\sum_{i=1}^{N_R}h^H_{k_\nu,i}\upsilon_i+ z_{k_\nu},
\label{pist13}
\end{align}
for
\begin{equation}\label{pist14}
a_{k_\nu,\ell}(\bp)\triangleq \sum_{i: (\nu,\ell)\in\clN_i}\sum_{k'=1}^{|\clK_\nu|}\bp^i_{k'_\nu,\ell}h^H_{k_\nu,i}G^i_{k'_\nu,\ell},
\end{equation}
and
\begin{equation}\label{pist15}
b_{k_\nu,\ell,\nu'}(\bp)\triangleq \sum_{i: (\nu',\ell)\in\clN_i}\sum_{k'=1}^{|\clK_{\nu'}|}\bp^{i}_{k'_{\nu'},\ell}h^H_{k_{\nu},i}G^i_{k'_{\nu'},\ell},
\end{equation}
with
\[
\mathbb{E}\left\{|h^H_{k_\nu,i}\upsilon_i|^2\right\}=\epsilon h^H_{k_\nu,\ell}\left(\left[\Delta(\bx^i)\right]^2+I_{n_t}\right)h_{k_\nu,\ell}.
\]
Upon employing successive interference cancelation (SIC), user $k_\nu$ subtracts the detected and remodulated signal of
\begin{equation}\label{piexp5a}
\ds\sum_{\ell'=1}^{\ell-1}a_{k_\nu,\ell'}(\bp)s_{\nu,\ell'}
\end{equation}
from the RHS of (\ref{pist13}), yielding
\begin{eqnarray}\label{piexp5b}
\ds\sum_{\ell'=\ell}^La_{k_\nu,\ell'}(\bp)s_{\nu,\ell'}
+\sum_{\nu'\in\clN_{req}\setminus\{\nu\}}\sum_{\ell=1}^Lb_{k_\nu,\ell,\nu'}(\bp)s_{\nu',\ell}\nonumber\\
+\ds\sum_{i=1}^{N_R}h^H_{k_\nu,i}\upsilon_i+ z_{k_\nu}
\end{eqnarray}
for detecting $s_{\nu,\ell}$ by considering
\begin{equation}\label{piexp5c}
a_{k_\nu,\ell}(\bp)s_{\nu,\ell}
\end{equation}
in (\ref{piexp5b}) as the signal of interest, and
\begin{eqnarray}\label{piexp5c}
\ds\sum_{\ell'=\ell+1}^La_{k_\nu,\ell'}(\bp)s_{\nu,\ell'}
+\sum_{\nu'\in\clN_{req}\setminus\{\nu\}}\sum_{\ell=1}^Lb_{k_\nu,\ell,\nu'}(\bp)s_{\nu',\ell}\nonumber\\
+\ds\sum_{i=1}^{N_R}h^H_{k_\nu,i}\upsilon_i+ z_{k_\nu}
\end{eqnarray}
as the interference-plus-noise term. As such, the throughput of $s_{f_{\nu},\ell}$ at user $k_\nu$ is formulated as
\begin{equation}\label{pist19}
r_{k_\nu,\ell}(\bp,\bx)\triangleq \ln\left( 1+
\frac{|a_{k_\nu,\ell}(\bp)|^2}{\psi_{k_\nu,\ell}(\bp,\bx)}\right),
\end{equation}
for
\begin{align}\label{pist20}
&\psi_{k_\nu,\ell}(\bp,\bx)\nonumber\\
&\quad\triangleq \sum_{\ell'=\ell+1}^L|a_{k_\nu,\ell'}(\bp)|^2
+\ds \sum_{\nu'\in\clN_{req}\setminus\{\nu\}}\sum_{\ell=1}^L|b_{k_\nu,\ell,\nu'}(\bp)|^2\nonumber\\
&\qquad+\epsilon\sum_{i=1}^{N_R}h^H_{k_\nu,i}\left(\left[\Delta(\bx^i)\right]^2+I_{n_t}\right)h_{k_\nu,i}+\sigma,
\end{align}
which is a convex quadratic function.

The throughput of $s_{\nu,\ell}$ after SIC becomes:
\begin{equation}\label{pist17}
r_{\nu,\ell}(\bp,\bx)\triangleq \min_{k=1,\dots, |\clK_\nu|}r_{k_\nu,\ell}(\bp,\bx).
\end{equation}
The energy efficiency optimization related problem may then be formulated as\footnote{The final result should be divided by $\log_2(e)$ to express the energy efficiency in terms of bps/Hz/W}:
\begin{subequations}\label{pist21}
	\begin{align}
	\max_{\bp,\bx}\ \theta(\bp,\bx)\triangleq \min_{\nu=1,\dots, N_{req}}\frac{\sum_{\ell=1}^L r_{\nu,\ell}(\bp,\bx)}
	{\pi_\nu(\bp,\bx)}\nonumber\\
    \mbox{s.t.}\quad (\ref{pist8}),\label{pist21a}\\
	r_{\min}\leq r_{\nu,\ell}(\bp,\bx), (\nu,\ell)\in\clN_{req}\times \clL,\label{pist21b}\\
	\sum_{\nu=1}^{N_{req}}\pi^i_{\nu}(\bp,\bx) \leq P, i\in\clN_R,\label{pist21c}
	\end{align}
\end{subequations}
where $P$ is the maximum power budget of each RRH, and $r_{\min}$ is the minimum rate required for guaranteeing the delivery of all files requested within a certain delivery deadline \cite{AD14}. As such (\ref{pist21b}) constitutes the rate constraint to ensure a high-probability successful delivery according to Shannonian information theory.

\subsection{PGS Computation}
Note that this problem is non-convex because the objective function (OF) in (\ref{pist21a}) is not only non-smooth but also
non-concave, while both the SFT constraint (\ref{pist8}) and the rate-constraint (\ref{pist21b}) are non-convex. We now develop
a path-following algorithm, which iterates for finding ever better feasible points for (\ref{pist21}) with the aid of convex solvers. To this end, we have to derive $(i)$ an inner convex approximation for the SFT constraint (\ref{pist8}) and the rate-constraint (\ref{pist21b}); and $(ii)$ a lower-bounding concave approximation for the OF in (\ref{pist21a}).

Let $(\pk,\xk)$ be the specific feasible point for (\ref{pist21}) that is found by the $\kappa$th
iteration.
\subsubsection{Inner convex approximation for the SFT constraint (\ref{pist8})}
Using the inequality (\ref{fund7}) in the appendix yields
\begin{align}
&m_i(\bp,\bx^i)\nonumber\\
&\leq m_i(\pk,\xik)\!+\!\epsilon\la \left(\!\epsilon I_{n_t}\!+\!L_i(\pk)
\!+\!\epsilon\!\left[\Delta(\xik)\right]^2\right)^{\!-1}\ra\nonumber\\
&\quad +\la \left[\Delta(\xik)\right]^2\ra - \la \left( \left[\Delta(\xik)\right]^2\!+\!I_{n_t}\right)^{-1}\ra\nonumber\\
&\quad +\la \left(\epsilon I_{n_t}\!+\!L_i(\pk)\!+\!\epsilon\!\left[\Delta(\xik)\right]^2\right)^{-1}, \nonumber\\
&\qquad\ L_i(\bp)\!+\!\epsilon\!\left[\Delta(\bx^i)\right]^2\ra - 2\la \Delta(\xik),\Delta(\bx^i)\ra\nonumber\\
&\quad +\la I_{n_t}\!-\!\left( \left[\Delta(\xik)\right]^2\!+\!I_{n_t}\right)^{-1},\left[\Delta(\bx^i)\right]^2\ra
\label{pist22a}\\
&=\mk_i +\!\sum_{(\nu,\ell)\in\Xi^i}||\left(\epsilon I_{n_t}\!+\!L_i(\pk)\!+\!\epsilon\!\left[\Delta(\xik)\right]^2\right)^{-1/2}\nonumber\\
&\qquad\ \times\!\sum_{k=1}^{|\clK_\nu|}\bp^{i}_{k_\nu,\ell}G^i_{k_\nu,\ell}||^2\nonumber\\
&\quad +\epsilon||\left(\epsilon I_{n_t}\!+\!L_i(\pk)\!+\!\epsilon\left[\Delta(\xik)\right]^2\right)^{-1/2}\!\Delta(\bx^i)||^2\nonumber\\
&\quad -2\la \Delta(\xik),\Delta(\bx^i)\ra\nonumber\\
&\quad +||\left(I_{n_t}\!-\!\left( \left[\Delta(\xik)\right]^2\!+\!I_{n_t}\right)^{-1} \right)^{-1/2}\!\Delta(\bx^i)||^2\label{pist22b}\\
&\triangleq m_i^{(\kappa)}(\bp,\bx^i),\label{pist22}
\end{align}
for
\begin{align}
\mk_i &\triangleq m_i(\pk,\xik)\nonumber\\
&\quad + \epsilon\la \left(\epsilon I_{n_t}+L_i(\pk) + \epsilon\left[\Delta\xik)\right]^2\right)^{-1}  \ra\nonumber\\
&\quad +\la \left[\Delta\xik)\right]^2\ra
-\la \left( \left[\Delta(\xik)\right]^2+I_{n_t}\right)^{-1}\ra.\nonumber
\end{align}
The function $m_i^{(\kappa)}(\bp,\bx^i)$ is convex quadratic. The non-convex constraint (\ref{pist8}) seen
in (\ref{pist21}) is then innerly approximated by the following convex constraint:
\begin{equation}
m_i^{(\kappa)}(\bp,\bx^i) \leq \log_2(e)C, i\in\clN_R. \label{pist23}
\end{equation}
\subsubsection{Inner convex quadratic approximation for the rate-constraint (\ref{pist21b})}
Furthermore, applying the inequality (\ref{fund5}) of the Appendix yields
\begin{align}
r_{k_\nu,\ell}(\bp,\bx)&\geq \ak_{k_\nu,\ell}
+2\Re\{\bk_{k_\nu,\ell}a_{k_\nu,\ell}(\bp)\}\nonumber\\
&\quad-\ck_{k_\nu,\ell}\left(|a_{k_\nu,\ell}(\bp)|^2 +\psi_{k_\nu,\ell}(\bp,\bx)\right)\label{pist24}\\
&\triangleq \rk_{k_\nu,\ell}(\bp,\bx)
\end{align}
in conjunction with
\[
\ak_{k_\nu,\ell}\triangleq r_{k_\nu,\ell}(\pk,\xk)-
\frac{|a_{k_\nu,\ell}(\pk)|^2}{\psi_{k_\nu,\ell}(\pk,\xk)},
\]
and
\[
\bk_{k_\nu,\ell}\triangleq \frac{a^H_{k_\nu,\ell}(\pk)}{\psi_{k_\nu,\ell}(\pk,\xk)},
\]
as well as
\begin{align}
0< \ck_{k_\nu,\ell}\triangleq &\frac{|a_{k_\nu,\ell}(\pk)|^2}{ \psi_{k_\nu,\ell}(\pk,\xk)}\nonumber\\
&\times \frac{1}{(|a_{k_\nu,\ell}(\pk)|^2+\psi_{k_\nu,\ell}(\pk,\xk))}.\nonumber
\end{align}
Since the function $r_{k_\nu,\ell}^{(\kappa)}(\bp,\bx)$ is concave quadratic, the following
function provides a concave lower-bounding approximation for $r_{\nu,\ell}(\bp,\Delta)$:
\[
r_{\nu,\ell}^{(\kappa)}(\bp,\bx)\triangleq \min_{k=1,\dots, |\clK_\nu|}r_{k_\nu,\ell}^{(\kappa)}(\bp,\bx),
\]
leading to the following inner convex quadratic approximation for the non-convex constraint (\ref{pist21b}) in (\ref{pist21})
\begin{equation}
\rk_{\nu,\ell}(\bp,\bx) \geq r_{\min}, (\nu,\ell) \in \clN_{req} \times \clL. \label{pist25}
\end{equation}
\subsubsection{PGS Algorithm}
We now solve the following convex optimization problem at the $\kappa$-th iteration
for generating the next feasible point $(p^{(\kappa+1)},x^{(\kappa+1)})$ for (\ref{pist21})
\begin{eqnarray}\label{pist26}
\ds\max_{\bp,\bx}\min_{\nu=1,\dots, N_{req}}[\sum_{\ell=1}^L\rk_{\nu,\ell}(\bp,\bx)\!-\!
\theta(\pk,\xk)\pi_\nu(\bp,\bx)] \nonumber\\
	\mbox{s.t.} \quad (\ref{pist21c}), (\ref{pist23}), (\ref{pist25}),
\end{eqnarray}
which is equivalent to the following convex quadratic problem:
\[
\begin{array}{r}
\ds\max_{\bp,\bx,\gamma}\ \gamma\quad\mbox{s.t} \quad (\ref{pist21c}), (\ref{pist23}), (\ref{pist25}),\\
\ds\sum_{\ell=1}^L\rk_{\nu,\ell}(\bp,\bx)\!-\!\theta(\pk,\xk)\pi_\nu(\bp,\bx)\geq \gamma,
\nu=1,\dots, N_{req}.
\end{array}
\]
The computational complexity of (\ref{pist26}), which involves $n_v=\sum_{i=1}^{N_R}|\clN_i|+N_Rn_t+1$ decision variables and
		$n_c=2N_R+N_{req}(L+1)$ constraints is on the order of \cite[p.4]{peaucelle2002user}
	\begin{equation}\label{cc}
	{\cal O}\left\{(n_c)^{2.5}\left[(n_v)^2+n_c\right]\right\}.
	\end{equation}
As $(\pko,\xko)$ and $(\pk,\xk)$ are the optimal solution and a feasible point
 for (\ref{pist26}), we have
\[
\begin{array}{lll}
\ds\min_{\nu=1,\dots, N_{req}}[\sum_{\ell=1}^L\rk_{\nu,\ell}(\pko,\xko)\\
\qquad \qquad -\theta(\pk,\xk)\pi_\nu(\pko,\xko)]\!\!\!\!&>&\\
\ds\min_{\nu=1,\dots, N_{req}}[\sum_{\ell=1}^L\rk_{\nu,\ell}(\pk,\xk)\\
\qquad \qquad \qquad -\theta(\pk,\xk)\pi_\nu(\pk,\xk)]\!\!\!\!&=&\!\!\!\!0.
\end{array}
\]
Therefore, we have:
\begin{align}
&\sum_{\ell=1}^L\rk_{\nu,\ell}(\pko,\xko)\nonumber\\
&\quad -\theta(\pk,\xk)\pi_\nu(\pko,\xko)>0
\ \forall\ \nu=1,\dots, N_{req}\nonumber
\end{align}
that yields
\begin{equation}\label{thetak}
\theta(\pko,\xko)>\theta(\pk,\xk),
\end{equation}
i.e. $(\pko,\xko)$ provides a better feasible point for the nonconvex problem (\ref{pist21})
than $(\pk,\xk)$, provided that $(\pko,\xko)\neq (\pk,\xk)$. As such, the sequence $\{(\pk,\xk)\}$ converges  to a locally optimal solution of (\ref{pist21}) \cite{MW78}. However, in the scenario of \cite{Nasetal17tcom}, this locally optimal solution turned out to be the globally optimal one.
Algorithm \ref{palg1} provides  the pseudo-code of this computational procedure.

To find a feasible initial point for (\ref{pist21}),   initialized by any feasible point for the convex constraints (\ref{pist21c}), we iterate
\begin{eqnarray}\label{pist27}
\ds\max_{\bp,\bx}\ \min \left\{\min_{(\nu,\ell)\in\clN_{req}\times \clL}[\frac{\rk_{\nu,\ell}(\bp,\bx)}{r_{\min}/\log_2(e)}-1],\right.\nonumber\\
\left.\min_{i\in\clN_R}
[1-\frac{1}{\log_2(e)C}m_i^{(\kappa)}(\bp,\bx^i)] \right\}\nonumber\\
\mbox{s.t.}\quad (\ref{pist21c}),
\end{eqnarray}
until the value of the objective in (\ref{pist27}) becomes more than $0$ or equal to $0$.
The problem (\ref{pist27}) is
equivalent to the following convex quadratic problem:
\begin{align}
\ds\max_{\bp,\bx,\gamma}\ \gamma\quad \mbox{s.t.}\quad  (\ref{pist21c}),\nonumber\\
\ds \rk_{\nu,\ell}(\bp,\bx)\geq r_{\min}(1+\gamma)/\log_2(e), (\nu,\ell)\in\clN_{req}\times \clL,\nonumber\\
\ds  m_i^{(\kappa)}(\bp,\bx^i)\leq \log_2(e)C (1-\gamma), i\in\clN_R.\nonumber
\end{align}

\begin{algorithm}[!t]
	\caption{RZFB PGS Algorithm} \label{palg1}
	\begin{algorithmic}[1]
	\State  Set $\kappa=0$. Iterate (\ref{pist27}) for a feasible point $(p^{(0)}, x^{(0)})$ for
(\ref{pist21}).
		\State \textbf{Repeat until convergence.}
		 Solve the convex optimization problem (\ref{pist26}) to generate the next feasible point
$(p^{(\kappa+1)}, x^{(\kappa+1)})$ for (\ref{pist21}).  Set $\kappa := \kappa + 1$.
\State \textbf{Output} $(p^{(\kappa)},\xk)$ as the optimal solution of (\ref{pist21}).
	\end{algorithmic}
\end{algorithm}

{\bf Remark 1.} Instead of the energy efficiency optimization problem (\ref{pist21}) one can consider
the following  problem of max-min rate optimization
\begin{equation}\label{pist21m}
\max_{\bp,\bx}\ \min_{(\nu,\ell)\in\clN_{req}\times \clL} r_{\nu,\ell}(\bp,\bx)
\quad\mbox{s.t.}\quad  (\ref{pist8}), (\ref{pist21c}).
\end{equation}
Then, instead of (\ref{pist26}), the following problem is solved at the $\kappa$-th iteration to generate a better
feasible point $(p^{(\kappa+1)}, x^{(\kappa+1)})$:
\begin{equation}\label{pist26m}
\ds\max_{\bp,\bx}\min_{(\nu,\ell)\in\clN_{req}\times \clL}\rk_{\nu,\ell}(\bp,\bx)\quad
	\mbox{s.t.} \quad (\ref{pist21c}), (\ref{pist23}),
\end{equation}
while for an initial feasible point, we choose a reasonable $r_{\min}$ and then iterate (\ref{pist27}).

\section{RZFB based improper Gaussian signaling}
\subsection{IGS Problem statement}
Observe  that the `beamformed' signal $\lambda^i_{\nu,\ell}(s_{\nu,\ell})$
defined by (\ref{prec2}) is proper Gaussian
($\mathbb{E}((\lambda^i_{\nu,\ell}(s_{\nu,\ell}))^2)=0$), because so is the information source
$s_{\nu,\ell}\in{\cal CN}(0,1)$. As such,
the RZFB technique of the previous section represents proper Gaussian signaling (PGS).

In this section, we propose the following new class of RZFB
\begin{align}
\lambda^i_{\nu,\ell}(s_{\nu,\ell})&=\sum_{k=1}^{|\clK_{\nu}|}\left(\bp^{i,1}_{k_\nu,\ell}G^i_{k_\nu,\ell}s_{\nu,\ell}+
\bp^{i,2}_{k_\nu,\ell}G^i_{k_\nu,\ell}s^*_{\nu,\ell}\right) \label{rec2a}\\
&=\left(\sum_{k=1}^{|\clK_{\nu}|}\bp^{i,1}_{k_\nu,\ell}G^i_{k_\nu,\ell}\right)\!s_{\nu,\ell}\!+\!
\left(\sum_{k=1}^{|\clK_{\nu}|}\bp^{i,2}_{k_\nu,\ell}G^i_{k_\nu,\ell}\right)\!s^*_{\nu,\ell}, \label{rec2}\nonumber\\
&\quad\ \bp^i_{k_\nu,\ell}\triangleq (\bp^{i,1}_{k_\nu,\ell},\bp^{i,2}_{k_\nu,\ell})\in\mathbb{C}^2,
\end{align}
where $\lambda^i_{\nu,\ell}(s_{\nu,\ell})$ is an improper Gaussian signal ($\mathbb{E}((\lambda^i_{\nu,\ell}(s_{\nu,\ell})^2)\neq 0$) capable of offering additional degree of freedom for improving the throughput of the proper Gaussian information source $s_{\nu,\ell}$ of section III \cite{SS10}. Explicitly, this is regarded as improper Gaussian signaling (IGS) in contrast to the proper Gaussian signaling (PGS) scheme of (\ref{prec2}), which has been shown to enhance the throughput of severely interference-limited networks
\cite{Tuaetal19,Yuetal20tcom,Naetal20tcom} . The reader is also referred to \cite{Yuetal21tvt} for another instantiation of IGS based (traditional) RZFB.

Based on the IGS philosophy of (\ref{rec2}), we rewrite (\ref{ist3a}) in an form as:
\begin{equation}\label{ist6}
\xi^A_i\triangleq \begin{bmatrix} \xi_i\cr
\xi_i^*\end{bmatrix}
=\sum_{(\nu,\ell)\in\Xi^i}\left(\sum_{k=1}^{|\clK_\nu|}\clG^i_{k_\nu,\ell}(\bp^{i}_{k_\nu,\ell})\right)
s^A_{\nu,\ell}+\upsilon^A_i,
\end{equation}
for
\begin{align}\label{ist6a}
&s^A_{\nu,\ell}=\begin{bmatrix}s_{\nu,\ell}\cr
s^*_{\nu,\ell}\end{bmatrix},
\upsilon^A_i=\begin{bmatrix}\upsilon_i\cr
\upsilon_i^*\end{bmatrix},\nonumber\\
&\clG^i_{k_\nu,\ell}(\bp^{i}_{k_\nu,\ell})\nonumber\\
&\quad\triangleq \begin{bmatrix}\bp^{i,1}_{k_\nu,\ell}G^i_{k_\nu,\ell}&
\bp^{i,2}_{k_\nu,\ell}G^i_{k_\nu,\ell}\\
(\bp^{i,2}_{k_\nu,\ell})^*(G^i_{k_\nu,\ell})^*&(\bp^{i,1}_{k_\nu,\ell})^*(G^i_{k_\nu,\ell})^*\end{bmatrix}
\in\mathbb{C}^{(2n_t)\times 2}.
\end{align}
For $\Delta^A(\tbx^i)\triangleq \mbox{diag}[\Delta(\tbx^i),\Delta(\tbx^i)]\in\mathbb{R}^{(2n_t)\times (2n_t)}$,
$\bp^i\triangleq \{\bp^i_{k_\nu,\ell}, (\nu,\ell)\in\clN_i, i=1,\dots, N_R, k_{\nu}=1,\dots, |\clK_{\nu}|\}$, and
\[
\Lambda_i(\bp^i)\triangleq \sum_{(\nu,\ell)\in\Xi^i}\left[\sum_{k=1}^{|\clK_{\nu}|}\clG^i_{\nu,\ell}(\bp^{i}_{\nu,\ell})\right]^2\in\mathbb{C}^{(2n_t)\times (2n_t)},
\]
the mutual information between $\sum_{(\nu,\ell)\in\Xi^i}\lambda^i_{\nu,\ell}(s_{\nu,\ell})$ and
quantized version $\xi$ is \cite{CT06}
\begin{align}
&\frac{1}{2}{\cal I}\left(\xi^A_i,\sum_{(\nu,\ell)\in\Xi^i}\left(\sum_{k=1}^{|\clK_\nu|}\clG^i_{k_\nu,\ell}(\bp^{i}_{k_\nu,\ell})\right)
s^A_{\nu,\ell}\right)\nonumber\\
&\quad=\frac{1}{2}\ln\left|I_{2n_t}+\Lambda_i(\bp^i)\left(\left[\Delta^A(\tbx^i)\right]^2+\epsilon I_{2n_t}\right)^{-1}\right|\nonumber\\
&\quad=\frac{1}{2}\ln\left|I_{2n_t}+\epsilon^{-1}\Lambda_i(\bp^i)\left(\left[\Delta^A(\bx^i)\right]^2+I_{2n_t}\right)^{-1}\right|\nonumber\\
&\quad\triangleq \mu_i(\bp,\bx^i),\label{ist7}
\end{align}
for
\begin{equation}\label{bx}
\bx^i=\tbx_i/\sqrt{\epsilon}.
\end{equation}
Similarly to the PGS expression of (\ref{pist8}), the fronthaul-rate constraint (\ref{mut}) is reformulated as:
\begin{equation}\label{ist8}
\mu_i(\bp,\bx^i)\leq \log_2(e)C, i\in\clN_R.
\end{equation}
Instead of (\ref{pist8a}), the signal transmitted by RRH $i$ now becomes
\begin{equation}\label{ist8b}
\chi_i=\sum_{(\nu,\ell)\in \clN_i}\left(\sum_{k=1}^{|\clK_\nu|}\begin{bmatrix}\bp^{i,1}_{k_\nu,\ell}G^i_{k_\nu,\ell}
&\bp^{i,2}_{k_\nu,\ell}G^i_{k_\nu,\ell}\end{bmatrix}\right)s^A_{\nu,\ell}+\upsilon_i.
\end{equation}
Instead of (\ref{pist8b}), the power consumption of transmitting the signal carrying $f_{\nu}$ from RRH $i$ is
\begin{eqnarray}
\frac{1}{2}\!\sum_{i=1}^{N_R}\!\sum_{\ell: (\nu,\ell)\in\clN_i}\!\!||\!\sum_{k=1}^{|\clK_{\nu}|}\clG^i_{k_\nu,\ell}(\bp^{i}_{k_\nu,\ell})||^2\!+\!
\sum_{i\in\iota_{\nu}}\!\sum_{j=1}^{n_t}\!\left[(\tbx^i_j)^2\!+\!\epsilon\right]\!\!=\!\!\! &&\nonumber\\
\frac{1}{2}\!\sum_{i=1}^{N_R}\!\sum_{\ell: (\nu,\ell)\in\clN_i}\!\!||\!\sum_{k=1}^{|\clK_{\nu}|}\clG^i_{k_\nu,\ell}(\bp^{i}_{k_\nu,\ell})||^2\!+\!
\epsilon\!\sum_{i\in\iota_{\nu}}\!\sum_{j=1}^{n_t}\!\left[(\bx^i_j)^2\!+\!1\right]\!\!\triangleq\!\!\! &&\nonumber\\
\tpi^i_\nu(\bp,\bx),\label{ist8b}&&
\end{eqnarray}
for $\bx\triangleq \{\bx^i, i=1,\dots, N_R\}$,
where $\iota_{\nu}$ is defined from (\ref{ist8a}) as  the set of RRHs that do not have the
entire file $\nu$ in their cache.

Instead of (\ref{pist9}), the total power consumption of delivering $f_{\nu}$ becomes
\begin{equation}\label{ist9}
\tpi_{\nu}(\bp,\bx)\triangleq \sum_{i=1}^{N_R}\tpi^i_{\nu}(\bp,\bx) +P_{non}.
\end{equation}
Under the RZFB (\ref{rec2}), the augmented form of equation (\ref{rec}) formulating the signal received by user $k_\nu$ is
\begin{align}
y^A_{k_\nu}&\triangleq\begin{bmatrix}y_{k_\nu}\cr
y^*_{k_\nu}\end{bmatrix}\label{ist11}\\
&=\ds\sum_{\ell=1}^L\clA_{k_\nu,\ell}(\bp)s^A_{\nu,\ell}
+\sum_{\nu'\in\clN_{req}\setminus\{\nu\}}\sum_{\ell=1}^L\clB_{k_\nu,\ell,\nu'}(\bp)s^A_{\nu',\ell}\nonumber\\
&\quad+\ds\sum_{i=1}^{N_R}\clH_{k_\nu,i}\upsilon^A_i+ z^A_{k_\nu},
\label{ist13}
\end{align}
for
\begin{equation}\label{ist16}
\clH_{k_\nu,i}\triangleq
\begin{bmatrix}h_{k_\nu,i}^H&0_{1\times n_t}\cr
0_{1\times n_t}&h_{k_\nu,i}^T\end{bmatrix}\in\mathbb{C}^{2\times (2n_t)}, z^A_{k_\nu}\triangleq
\begin{bmatrix}z_{k_\nu}\cr
z^*_{k_\nu}
\end{bmatrix}
\end{equation}
and
\begin{equation}\label{ist14}
\clA_{k_\nu,\ell}(\bp)\triangleq \sum_{i: (\nu,\ell)\in\clN_i}\clH_{k_\nu,i}\sum_{k'=1}^{|\clK_\nu|}\clG^i_{k'_\nu,\ell}(\bp^i_{k'_\nu,\ell})
\in\mathbb{C}^{2\times 2},
\end{equation}
as well as
\begin{equation}\label{ist15}
\clB_{k_\nu,\ell,\nu'}(\bp)\triangleq \sum_{i: (\nu',\ell)\in\clN_i}\clH_{k_{\nu},i}\sum_{k'=1}^{|\clK_{\nu'}|}\clG^i_{k'_{\nu'},\ell}(\bp^{i}_{k'_{\nu'},\ell})
\in\mathbb{C}^{2\times 2}.
\end{equation}
It is readily shown that
\[
\mathbb{E}\left\{[\clH_{k_\nu,i}^H\upsilon^A_i]^2\right\}=\epsilon \clH_{k_\nu,\ell}\left(\left[\Delta^A(\bx^i)\right]^2+I_{2n_t}\right)\clH_{k_\nu,\ell}^H.
\]
In SIC, user $k_\nu$ subtracts the detected and remodulated signal
\begin{equation}\label{iexp5a}
\ds\sum_{\ell'=1}^{\ell-1}\clA_{k_\nu,\ell'}(\bp)s^A_{\nu,\ell'}
\end{equation}
from the RHS of (\ref{ist13}), yielding
\begin{eqnarray}\label{iexp5b}
\ds\sum_{\ell'=\ell}^L\clA_{k_\nu,\ell'}(\bp)s^A_{\nu,\ell'}
+\sum_{\nu'\in\clN_{req}\setminus\{\nu\}}\sum_{\ell=1}^L\clB_{k_\nu,\ell,\nu'}(\bp)s^A_{\nu',\ell}\nonumber\\
+\ds\sum_{i=1}^{N_R}\clH_{k_\nu,i}\upsilon^A_i+ z^A_{k_\nu}
\end{eqnarray}
for detecting $s_{\nu,\ell}$ by considering
\begin{equation}\label{iexp5c}
\clA_{k_\nu,\ell}(\bp)s^A_{\nu,\ell}
\end{equation}
in (\ref{iexp5b}) as the signal of interest, and
\begin{eqnarray}\label{iexp5c}
\ds\sum_{\ell'=\ell+1}^L\clA_{k_\nu,\ell'}(\bp)s^A_{\nu,\ell'}
+\sum_{\nu'\in\clN_{req}\setminus\{\nu\}}\sum_{\ell=1}^L\clB_{k_\nu,\ell,\nu'}(\bp)s^A_{\nu',\ell}\nonumber\\
+\ds\sum_{i=1}^{N_R}\clH_{k_\nu,i}\upsilon^A_i+ z^A_{k_\nu}
\end{eqnarray}
as the interference-plus-noise. As such the throughput $s_{f_{\nu},\ell}$ of user $k_\nu$ is calculated as \cite{T99}
\begin{align}\label{iexp5d}
\frac{1}{2}{\cal I}&\left(\!\clA_{k_\nu,\ell}(\bp)s^A_{\nu,\ell},\ds\sum_{\ell'=\ell+1}^L\clA_{k_\nu,\ell'}(\bp)s^A_{\nu,\ell'}\right.\nonumber\\
&\left.\!+\!\sum_{\nu'\in\clN_{req}\setminus\{\nu\}}\sum_{\ell=1}^L\clB_{k_\nu,\ell,\nu'}(\bp)s^A_{\nu',\ell}
\!+\!\ds\sum_{i=1}^{N_R}\clH_{k_\nu,i}\upsilon_i\!+\!z_{k_\nu}\!\right)\!.
\end{align}
Then, based on \cite{SS10} we have:
\begin{equation}\label{ist19}
\rho_{k_\nu,\ell}(\bp,\bx)\triangleq \ln\left| I_2+
[\clA_{k_\nu,\ell}(\bp)]^2\Psi^{-1}_{k_\nu,\ell}(\bp,\bx)\right|,
\end{equation}
where
\begin{align}\label{ist20}
&\Psi_{k_\nu,\ell}(\bp,\bx)\nonumber\\
&\quad \triangleq \sum_{\ell'=\ell+1}^L[\clA_{k_\nu,\ell'}(\bp)]^2
+\ds \sum_{\nu'\in\clN_{req}\setminus\{\nu\}}\sum_{\ell=1}^L[\clB_{k_\nu,\ell,\nu'}(\bp)]^2\nonumber\\
&\qquad+\epsilon\sum_{i=1}^{N_R}\clH_{k_\nu,i}\left(\left[\Delta^A(\bx^i)\right]^2+I_{2n_t}\right) \clH_{k_\nu,i}^H+\sigma I_2.
\end{align}
The throughput of $s_{\nu,\ell}$ attained by SIC thus becomes:
\begin{equation}\label{ist17}
\frac{1}{2}\rho_{\nu,\ell}(\bp,\bx),
\end{equation}
where we have:
\begin{equation}\label{ist18}
\rho_{\nu,\ell}(\bp,\bx)=\min_{k=1,\dots, |\clK_\nu|}\rho_{k_\nu,\ell}(\bp,\bx).
\end{equation}

The problem of IGS energy efficiency optimization is reminiscent of its PGS counterpart in (\ref{pist21}),
which is formulated as\footnote{Again the final result should be divided by $\log_2(e)$ to convert the energy-efficiency to
bps/Hz/W}:
\begin{subequations}\label{ist21}
	\begin{align}
	\max_{\bp,\bx}\ \Theta(\bp,\bx)\triangleq \min_{\nu=1,\dots, N_{req}}\frac{\sum_{\ell=1}^L\rho_{\nu,\ell}(\bp,\bx)}
	{2\tpi_\nu(\bp,\bx)}\nonumber\\
    \mbox{s.t.}\quad (\ref{ist4}),(\ref{ist8}),\label{ist21a}\\
	2r_{\min}\leq \rho_{\nu,\ell}(\bp,\bx), (\nu,\ell)\in\clN_{req}\times \clL,\label{ist21b}\\
	\sum_{\nu=1}^{N_{req}}\tpi^i_{\nu}(\bp,\bx) \leq P, i\in\clN_R.\label{ist21c}
	\end{align}
\end{subequations}
This is non-convex, because the OF of (\ref{ist21a}) is both non-smooth and non-concave, while both the fronthaul-rate constraint (\ref{ist8}) and (\ref{ist21b}) are non-convex.
In contrast to the PGS throughput defined by (\ref{pist19}), which is a logarithmic function, the IGS throughput defined by (\ref{ist19}) is a log-determinant function. As a result, the computational algorithms developed in the previous section are not applicable to address the problem (\ref{ist21}).

\subsection{IGS Computation}
Similarly to PGS, we now derive an inner convex approximation for both (\ref{ist8}) as well as (\ref{ist21b}) and a concave lower-bounding approximation for the OF in (\ref{ist21a}) that are used for a path-following computational procedure to generate a sequence of gradually improved feasible points for (\ref{ist21}).

Let $(\pk,\xk)$ be the specific feasible point for (\ref{ist21}) that is found by the $\kappa$th
iteration.
\subsubsection{Inner convex quadratic approximation for the SFT constraint (\ref{ist8})}
Similarly to (\ref{pist22}), we use the inequality (\ref{fund7}) of the Appendix for obtaining the following
convex quadratic upper-bounding approximation for the function defined in (\ref{ist7}):
\begin{align}
&\mu_i(\bp,\bx^i)\nonumber\\
&\leq \muk_i\!+\!\sum_{(\nu,\ell)\in\Xi^i}||\left(\epsilon I_{n_t}\!+\!\Lambda_i(\pk)\!+\!\epsilon\left[\Delta(\xik)\right]^2\right)^{\!-1/2}\nonumber\\
&\qquad\ \times\!\sum_{k=1}^{|\clK_{\nu}|}\clG^i_{k_\nu,\ell}(\bp^{i}_{k_\nu,\ell}) ||^2\nonumber\\
&\quad+\!\epsilon||\left(\epsilon I_{2n_t}\!+\!\Lambda_i(\pk)\!+\!\epsilon\left[\Delta^A(\xik)\right]^2\right)^{\!-1/2}\!\Delta^A(\bx^i)||^2\nonumber\\
&\quad-\!2\la \Delta^A(\xik),\Delta^A(\bx^i)\ra\nonumber\\
&\quad+\!||\left(I_{2n_t}\!-\!\left( \left[\Delta^A(\xik)\right]^2\!+\!I_{2n_t}\right)^{-1} \right)^{\!-1/2}\!\Delta^A(\bx^i)||^2\label{ist22b}\\
&\triangleq \mu_i^{(\kappa)}(\bp,\bx^i),\label{ist22}
\end{align}
for
\begin{align}
\muk_i &\triangleq \mu_i(\pk,\xik)\nonumber\\
&\quad +\frac{1}{2}\left[\epsilon\la \left(\epsilon I_{2n_t}+\Lambda_i(\pik)
+\epsilon\left[\Delta^A(\xik)\right]^2\right)^{-1} \ra \right. \nonumber\\
&\quad \left.+\la \left[\Delta^A(\xik)\right]^2\ra - \la \left( \left[\Delta^A(\xik)\right]^2+I_{2n_t}\right)^{-1}\ra\right]\!.\nonumber
\end{align}
The non-convex constraint (\ref{ist8})
in (\ref{ist21}) is then innerly approximated by the following convex quadratic constraint:
\begin{equation}
\mu_i^{(\kappa)}(\bp,\bx^i) \leq \log_2(e)C, i\in\clN_R. \label{ist23}
\end{equation}
\subsubsection{Inner convex quadratic approximation for the rate-constraint (\ref{ist21b})}
Furthermore, applying the inequality (\ref{fund6}) of the Appendix yields
\begin{align}
&\rho_{k_\nu,\ell}(\bp,\bx)\nonumber\\
&\quad\geq \ak_{k_\nu,\ell}
+2\Re\{\la \Bk_{k_\nu,\ell}\clA_{k_\nu,\ell}(\bp)\ra\}\nonumber\\
&\qquad-\la \Ck_{k_\nu,\ell},[\clA_{k_\nu,\ell}(\bp)]^2+\Psi_{k_\nu,\ell}(\bp,\bx)\ra\label{ist24}\\
&\quad=\ak_{k_\nu,\ell}
+2\Re\{\la \Bk_{k_\nu,\ell}\clA_{k_\nu,\ell}(\bp)\ra\}-||(\Ck_{k_\nu,\ell})^{1/2}\clA_{k_\nu,\ell}(\bp)||^2\nonumber\\
&\qquad-\sum_{\ell'=\ell+1}^L||(\Ck_{k_\nu,\ell})^{1/2}\clA_{k_\nu,\ell'}(\bp)||^2\nonumber\\
&\qquad-\ds \sum_{\nu'\in\clN_{req}\setminus\{\nu\}}\sum_{\ell=1}^L||(\Ck_{k_\nu,\ell})^{1/2}\clB_{k_\nu,\ell,\nu'}(\bp)||^2\nonumber\\
&\qquad-\epsilon\sum_{i=1}^{N_R}||\left(\Ck_{k_\nu,\ell}\right)^{1/2}\clH_{k_\nu,i}\Delta^A(\bx^i)||^2\nonumber\\
&\qquad-\epsilon \sum_{i=1}^{N_R}||\left(\Ck_{k_\nu,\ell}\right)^{1/2}\clH_{k_\nu,i}||^2
-\sigma\la\Ck_{k_\nu,\ell}\ra,\nonumber\\
&\quad\triangleq \rhok_{k_\nu,\ell}(\bp,\bx),\label{ist24e}
\end{align}
with
\[
\ak_{k_\nu,\ell}\triangleq \rho_{k_\nu,\ell}(\pk,\xk)\!-\!
\la[\clA_{k_\nu,\ell}(\pk)]^2\Psi^{-1}_{k_\nu,\ell}(\pk,\xk)\ra,
\]
and
\[
\Bk_{k_\nu,\ell}\triangleq (\clA_{k_\nu,\ell}(\pk))^H\Psi^{-1}_{k_\nu,\ell}(\pk,\xk),
\]
as well as
\begin{align}
0\preceq \Ck_{k_\nu,\ell}\triangleq &\Psi^{-1}_{k_\nu,\ell}(\pk,\xk)\nonumber\\
&-\!\left([\clA_{k_\nu,\ell}(\pk)]^2\!+\!\Psi_{k_\nu,\ell}(\pk,\xk)\right)^{-1}.\nonumber
\end{align}
Since $\rhok_{k_\nu,\ell}(\bp,\bx)$ is a concave quadratic function, the
following function provides a lower-bounding concave approximation for $\rho_{\nu,\ell}(\bp,\Delta)$:
\[
\rho_{\nu,\ell}^{(\kappa)}(\bp,\bx)\triangleq \min_{k=1,\dots, |\clK_\nu|}\rho_{k_\nu,\ell}^{(\kappa)}(\bp,\bx),
\]
leading to the following inner convex quadratic approximation for the non-convex constraint (\ref{ist21b}) in (\ref{ist21})
\begin{equation}
\rho_{\nu,\ell}^{(\kappa)}(\bp,\bx) \geq 2r_{\min}, (\nu,\ell) \in \clN_{req} \times \clL. \label{ist25}
\end{equation}

\subsubsection{IGS Algorithm}
We now solve the following convex  problem at the $\kappa$-th iteration
for generating the next feasible point $(p^{(\kappa+1)},x^{(\kappa+1)})$ for (\ref{ist21})
\begin{eqnarray}\label{ist26}
\ds\max_{\bp,\bx}\min_{\nu=1,\dots, N_{req}}[\sum_{\ell=1}^L\rho^{(\kappa)}_{\nu,\ell}(\bp,\bx)\!-\!2\Theta(\pk,\xk)\tpi_\nu(\bp,\bx)]\nonumber\\
\mbox{s.t.} \quad (\ref{ist4}), (\ref{ist21c}), (\ref{ist23}), (\ref{ist25}),
\end{eqnarray}
which is equivalent to the convex quadratic problem:
\[
\begin{array}{r}
\ds\max_{\bp,\bx,\gamma}\ \gamma\quad\mbox{s.t} \quad (\ref{ist4}), (\ref{ist21c}), (\ref{ist23}), (\ref{ist25}),\\
\ds\!\sum_{\ell=1}^L\rho^{(\kappa)}_{\nu,\ell}(\bp,\bx)\!-\!2\Theta(\pk,\xk)\tpi_\nu(\bp,\bx)\!\geq\! \gamma,
\nu\!=\!1,\dots, N_{req}.
\end{array}
\]
The computational complexity of  (\ref{ist26}) is given by (\ref{cc}) with $n_v=2\sum_{i=1}^{N_R}|\clN_i|+N_Rn_t+1$ and $n_c=2N_R+N_{req}(L+1)$.

Similarly to (\ref{thetak}), we can show that
\[
\Theta(\pko,\xko)>\Theta(\pk,\xk),
\]
provided that $(\pko,\xko)\neq (\pk,\xk)$, so similarly to Algorithm \ref{palg1}, Algorithm \ref{alg1} also generates a sequence
$\{(\pk,\xk)\}$ of improved feasible points for (\ref{ist21}) that converges to a locally optimal solution
of (\ref{ist21}).

To find an initial feasible point for (\ref{ist21}),   initialized by any feasible point for the convex constraints (\ref{ist4}), (\ref{ist21c}) we iterate the following optimization procedure
\begin{eqnarray}\label{ist27}
\ds\max_{\bp,\bx}\ \min \left\{\min_{(\nu,\ell)\in\clN_{req}\times \clL}\left[\frac{\rho_{\nu,\ell}^{(\kappa)}(\bp,\bx)}{2r_{\min}/\log_2(e)}-1\right],\right.\nonumber\\
\left.\min_{i\in\clN_R}
\left[1-\frac{1}{\log_2(e)C}\mu_i^{(\kappa)}(\bp^i,\bx^i)\right] \right\}\nonumber\\
\mbox{s.t.}\quad (\ref{ist4}), (\ref{ist21c}),
\end{eqnarray}
until the value of the objective in (\ref{ist27}) becomes more than $0$ or equal to $0$. The problem (\ref{ist27}) is
equivalent to
\begin{align}
\ds\max_{\bp,\bx,\gamma}\ \gamma\quad \mbox{s.t.}\quad (\ref{ist4}), (\ref{ist21c}),\nonumber\\
\ds \rho_{\nu,\ell}^{(\kappa)}(\bp,\bx)\geq 2r_{\min}(1+\gamma)/\log_2(e), (\nu,\ell)\in\clN_{req}\times \clL,\nonumber\\
\ds  \mu_i^{(\kappa)}(\bp^i,\bx^i)\leq \log_2(e)C (1-\gamma), i\in\clN_R.\nonumber
\end{align}

\begin{algorithm}[!t]
	\caption{RZFB IGS Algorithm} \label{alg1}
	\begin{algorithmic}[1]
	\State  Set $\kappa=0$. Iterate (\ref{ist27}) for a feasible point $(p^{(0)}, x^{(0)})$ for
(\ref{ist21}).
		\State \textbf{Repeat until convergence.}
		 Solve the convex optimization problem (\ref{ist26}) to generate the next feasible point
$(p^{(\kappa+1)}, x^{(\kappa+1)})$ for (\ref{ist21}).  Set $\kappa := \kappa + 1$.
\State \textbf{Output} $(p^{(\kappa)},\xk)$ as the optimal solution of (\ref{ist21}).
	\end{algorithmic}
\end{algorithm}

Before closing this section, let us point out that Algorithm \ref{alg1} can also be readily adjusted to solve
the following max-min rate optimization problem:
\begin{equation}\label{ist21m}
\max_{\bp,\bx}\ \min_{(\nu,\ell)\in\clN_{req}\times \clL} \frac{1}{2}\rho_{\nu,\ell}(\bp,\bx)
\quad\mbox{s.t.}\quad (\ref{ist4}), (\ref{ist8}), (\ref{ist21c}).
\end{equation}
Then, instead of (\ref{ist26}) the following problem is solved at the $\kappa$-th iteration to generate
$(p^{(\kappa+1)}, x^{(\kappa+1)})$:
\begin{equation}\label{ist26m}
\ds\max_{\bp,\bx}\min_{(\nu,\ell)\in\clN_{req}\times \clL}\frac{1}{2}\rho^{(\kappa)}_{\nu,\ell}(\bp,\bx)\quad
	\mbox{s.t.} \quad (\ref{ist4}), (\ref{ist21c}), (\ref{ist23}),
\end{equation}
while for an initial feasible, we choose a reasonable $r_{\min}$ and then iterate (\ref{ist27}).

\begin{table}[!t]
	\centering
	\caption{Parameter Settings}
	\begin{tabular}{|l|l|}
		\hline
		Parameter & Value \\
		\hline
		Radius of cell   &  300 m  \\
		Carrier frequency/ Bandwidth(Bw)   &  2 GHz / 10 MHz   \\
		RRH maximum transmit power	&	30 dBm	\\
		Shadowing standard deviation &	8 dB \\
		Noise power density   &   $-174$ dBm/Hz   \\
		The circuit power per antenna ($P_a$)   &  5.6*1e-3 W   \\
        The number of transmit antennas at RRH ($n_t$)& 4\\
        The number of RRHs ($N_R$)&5\\
        The number of UEs  ($N_u$)&10\\
        The number of subfiles per file ($L$)&4\\
        The number of files in the library ($F$)&100\\
        Caching capacity ($\mu$)&1/2\\
        Popularity exponent for Zipf's distribution ($\gamma_z$)&1\\
        \hline
	\end{tabular}
	\label{table:1}
\end{table}

\section{Simulation Results} \label{sec:simulation}
The efficiency of the proposed algorithms along with their convergence is
demonstrated through the numerical results of this section.
The channel vector $h_{k,i}$ between the RRH $i$ and UE $k$ at the distance $d_{k,i}$ in km
is modeled as $h_{k,i} = \sqrt{10^{-\rho_{k,i}/10}} \tilde{h}_{k,i}$, where  $\rho_{k,i} = 148.1 + 37.6 \log_{10} (d_{k,i})$ (dB) is the pathloss and $\tilde{h}_{k,i}$ having independent and identically distributed complex entries is the small-scale fading  \cite{bjornson2013massive}. The error tolerance for declaring convergence is set to $\varepsilon =$ 1e-3.
Unless stated otherwise, the main parameters follow  Table \ref{table:1}.
We simulate three of the most popular caching strategies:
\begin{itemize}
	\item{ Caching the most popular files (CMP):}  Each RRH stores  $\clF_i$ number of the most popular files, so we have $|\clF_i|=\lfloor \mu F\rfloor$ and 	$c^i_{f,\ell}=1$ if and only if $f\leq |\clF_i|$.
	\item{Caching fractions of distinct files (CFD):}
	Each RRH stores up to $\lfloor  L/2\rfloor$ fragments of each file that are randomly chosen, so
	$	c^i_{f,\ell}=1$ if and only if $\ell\in\clL^i_f$,
where $\clL^i_f$ is a set of $\lfloor L/2\rfloor$ numbers picked randomly from $\clL\triangleq \{1,\dots, L\}$,
	which are independent across the file $f$ and the RRH index $i$.
	This strategy promotes collaboration between RRHs, when their cache capacity $C$ is small.
	\item{Random caching (RanC):}
	Each RRH stores a set $\clF_i$ number of distinct files, which are arbitrarily selected from $F$ files,
so $|\clF_i|=\lfloor \mu F\rfloor$ and
$	c^i_{f,l}= 1$ if and only  if $f\in \clF_i$.
\end{itemize}
Each UE is served by a single file at each time. The UE's requests are generated by Zipf's distribution associated with $\gamma_z=1$.

\begin{figure}[!t]
	\centerline{\includegraphics[width=8cm]{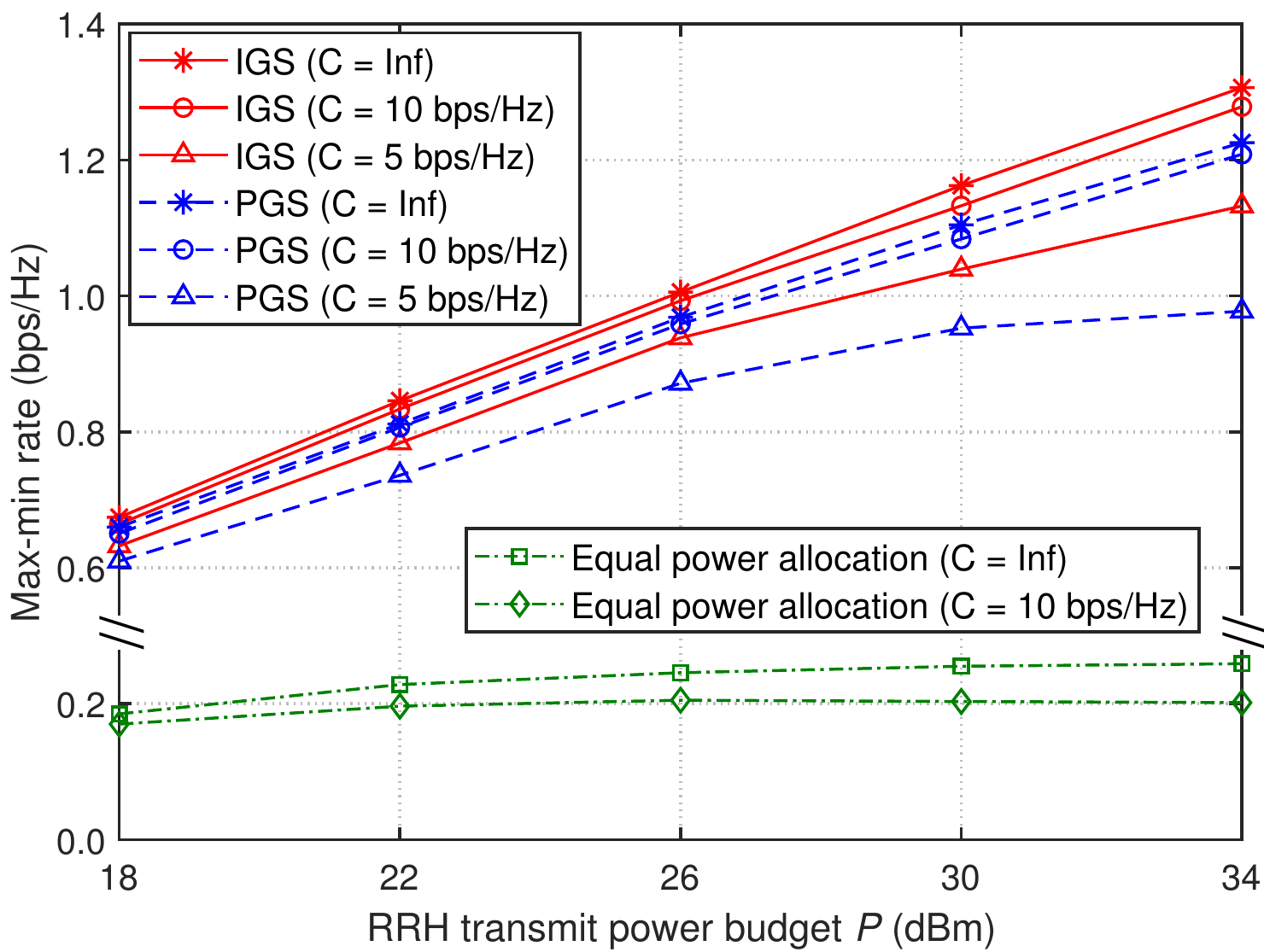}}
	\caption{The max-min rate versus RRH transmit power budget $P$ using the parameters of Table \ref{table:1}}
	\label{fig:rate_vs_P}
\end{figure}

\begin{figure}[!t]
	\centerline{\includegraphics[width=8cm]{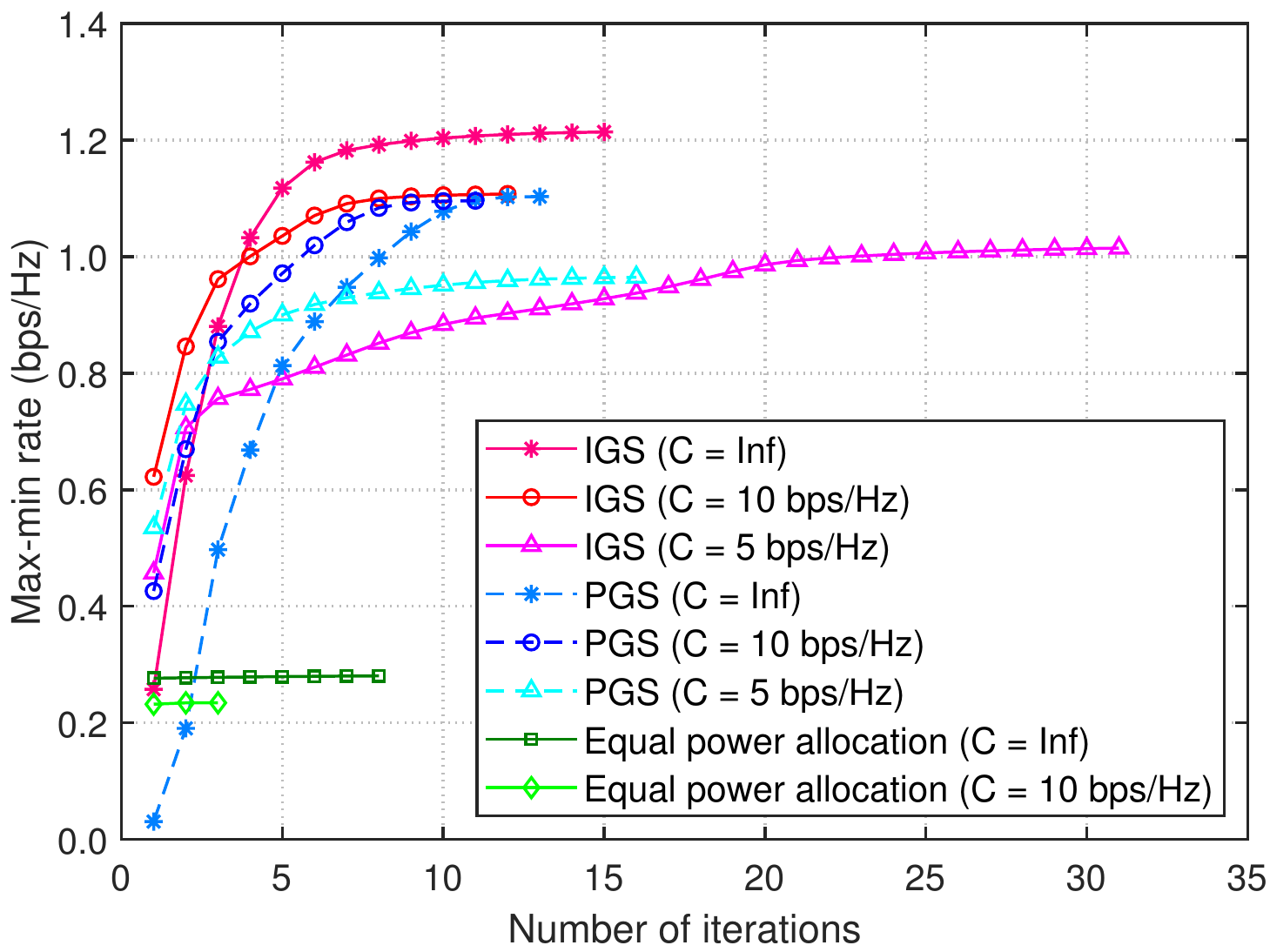}}
	\caption{The convergence behaviors of the max-min rate optimization schemes}
	\label{fig:conv_rate_vs_P}
\end{figure}

\begin{figure}[!t]
	\centerline{\includegraphics[width=8cm]{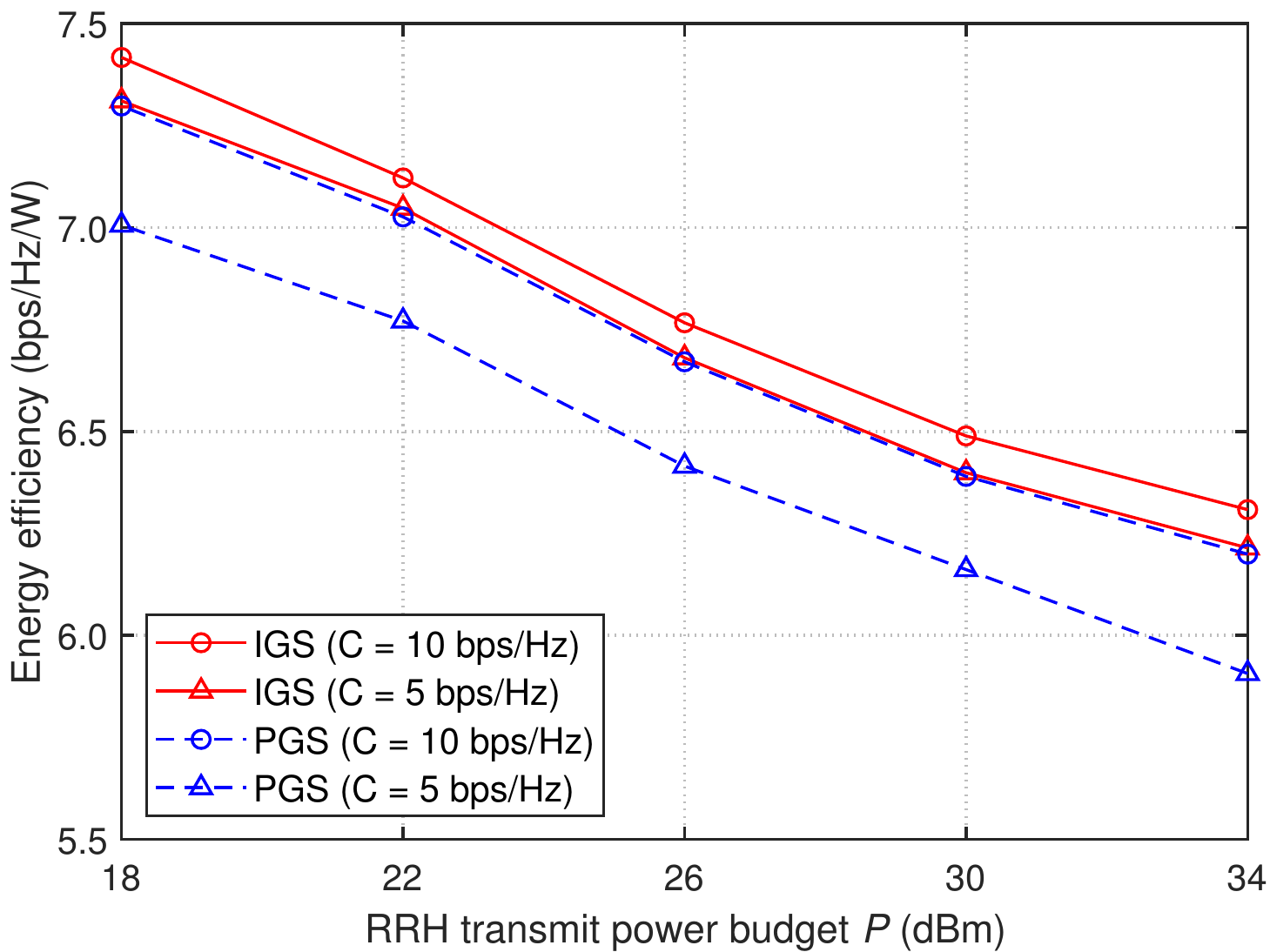}}
	\caption{The energy efficiency versus RRH transmit power budget $P$ using the parameters of Table \ref{table:1}}
	\label{fig:EE_vs_P}
\end{figure}

\begin{table*}[!t]
	\centering
	\caption{The average number of iterations for implementing IGS, PGS and equal power allocation algorithms in obtaining Fig. \ref{fig:rate_vs_P}}
	\begin{tabular}{|l|c|c|c|c|c|}
		\hline
		                      & $P = 18$ dBm & $P = 22$ dBm & $P = 26$ dBm & $P = 30$ dBm & $P = 34$ dBm \\ \hline
        IGS ($C =$ Inf)       & 9.6        & 8.9         & 12.0        & 16.3        & 20.3        \\
        IGS ($C = 10$ bps/Hz) & 10.3       & 9.0         & 14.1        & 13.6        & 17.4        \\
        IGS ($C = 5$ bps/Hz)  & 11.6       & 14.3        & 19.8        & 27.0        & 26.4        \\
        PGS ($C =$ Inf)       & 6.9        & 11.0        & 13.6        & 13.6        & 19.3        \\
        PGS ($C = 10$ bps/Hz) & 6.1        & 6.1         & 7.5         & 10.9        & 13.3        \\
        PGS ($C = 5$ bps/Hz)  & 10.4       & 17.4        & 14.0        & 13.6        & 18.8        \\
        Equal power allocation ($C =$ Inf)       & 2.0        & 2.6        & 3.5        & 6.3        & 4.9        \\
        Equal power allocation ($C = 10$ bps/Hz) & 4.0        & 5.6        & 5.3        & 5.3        & 5.1        \\ \hline
	\end{tabular}
	\label{table:2}
\end{table*}

\begin{table*}[!t]
	\centering
	\caption{The average number of iterations for implementing IGS and PGS algorithms in obtaining Fig. \ref{fig:EE_vs_P}}
	\begin{tabular}{|l|c|c|c|c|c|}
		\hline
                              & $P = 18$ dBm & $P = 22$ dBm & $P = 26$ dBm & $P = 30$ dBm & $P = 34$ dBm \\ \hline
        IGS ($C = 10$ bps/Hz) & 5.5         & 6.4         & 7.6         & 15.9        & 16.3        \\
        IGS ($C = 5$ bps/Hz)  & 9.0         & 9.8         & 11.9        & 14.1        & 15.9        \\
        PGS ($C = 10$ bps/Hz) & 4.6         & 6.3         & 6.8         & 8.1         & 12.0        \\
        PGS ($C = 5$ bps/Hz)  & 7.4         & 9.4         & 11.1        & 11.9        & 15.6        \\ \hline
	\end{tabular}
	\label{table:3}
\end{table*}

Fig. \ref{fig:rate_vs_P} plots the max-min user rate versus the transmit power budget $P$ at the RRH. Observe that as expected, the achievable max-min rate is improved upon increasing the transmit power budget. Furthermore, the IGS algorithm outperforms the PGS algorithm. Additionally, the max-min rate generated without any fronthaul constraint and that generated by observing the fronthaul constraint of either $C = 10$ bps/Hz or $C = 5$ bps/Hz are also illustrated in Fig. \ref{fig:rate_vs_P}. To expound further, observe that increasing the fronthaul capacity $C$ results in increased $r_{min}$ for both the IGS and PGS algorithms. The fixed power allocation policy is also characterized in Fig. \ref{fig:rate_vs_P}, in which the max-min rate optimization is carried out under the equal power allocation and uniform quantization. We can observe that the achievable max-min rate under this policy is much worse than that of the proposed RZFB. Table \ref{table:2} provides the average number of iterations required for declaring convergence in Fig. \ref{fig:rate_vs_P}.

Fig. \ref{fig:conv_rate_vs_P} plots the convergence behaviors of the max-min rate optimization schemes in simulating Fig. \ref{fig:rate_vs_P} with $ P = 30$ dBm. The max-min rate improves after each iteration by the proposed RZFB schemes, and the PGS algorithm converges faster than the IGS algorithm as the latter needs to handle more decision variables.

Fig. \ref{fig:EE_vs_P} plots the achievable energy efficiency (EE) versus the transmit power budget $P$ at the RRH with $r_{min}$ fixed at $0.4$ bps/Hz, where the fronthaul capacity $C$ set to either $10$ bps/Hz or $5$ bps/Hz. The proposed IGS algorithm outperforms the PGS algorithm in terms of its EE under different power budgets. As expected, a reduced EE is attained at an increased power budget. At the fronthaul capacity of $C=10$ bps/Hz, which allows the BBU to transfer the uncached contents to RRHs at higher rates, the EE improves for both the IGS and PGS algorithms. As the fronthaul capacity $C$ is reduced from $10$ bps/Hz to $5$ bps/Hz, the achievable EE of the IGS algorithm exhibits a modest but non-negligible performance erosion compared to that of PGS. The average number of iterations required for convergence in Fig. \ref{fig:EE_vs_P} is shown in Table \ref{table:3}.

\begin{figure}[!t]
	\centerline{\includegraphics[width=8cm]{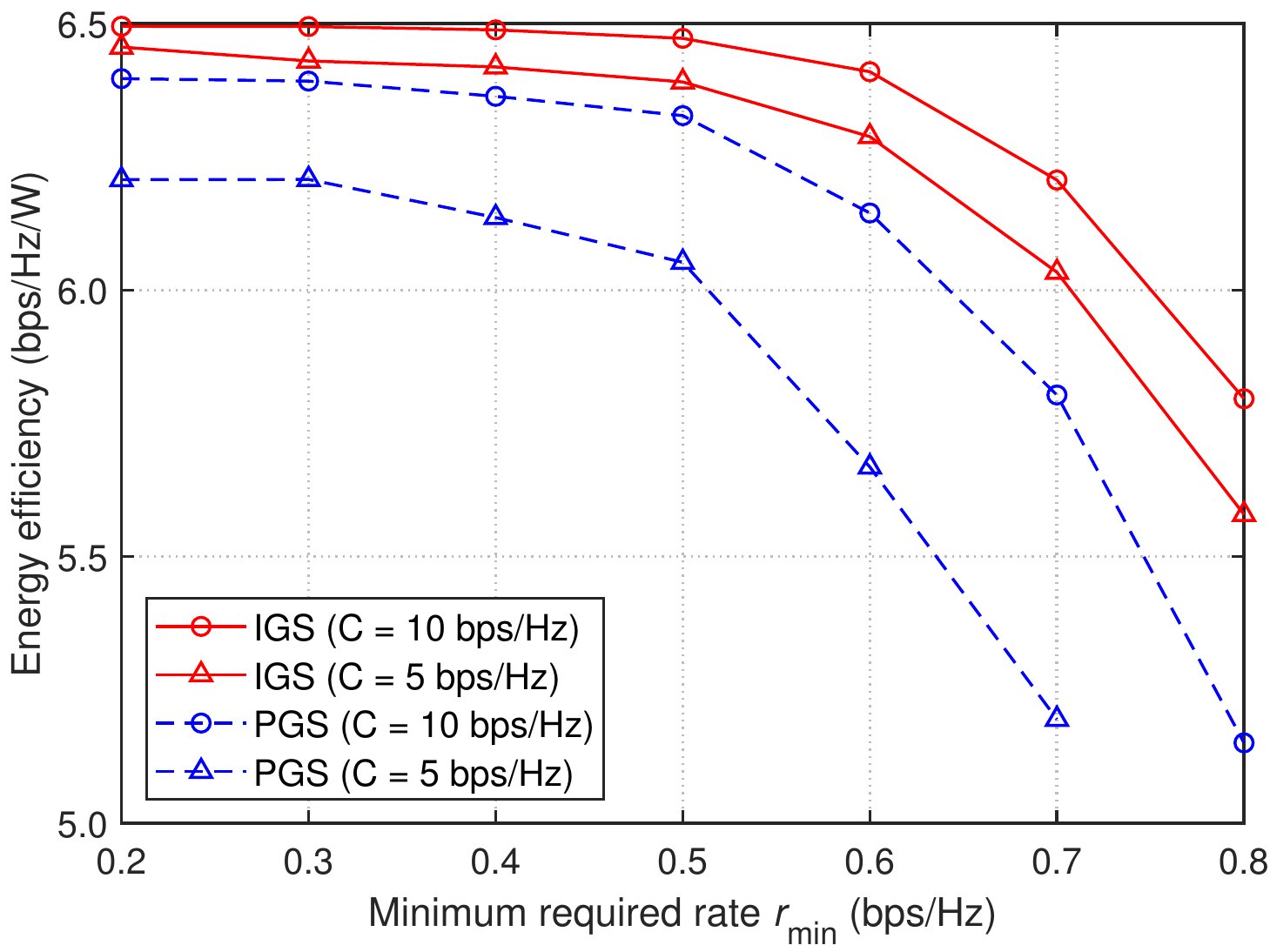}}
	\caption{The energy efficiency versus the minimum required rate $r_{min}$ using the parameters of Table \ref{table:1}}
	\label{fig:EE_vs_rmin}
\end{figure}

Fig. \ref{fig:EE_vs_rmin} plots the achievable EE versus the minimum required rate $r_{min}$ with the fronthaul capacity $C$ either set to $10$ bps/Hz or to $5$ bps/Hz. The transmit power budget $P$ is fixed at $30$ dBm in the following simulations. Observe that the EE of IGS algorithm is higher than that of PGS. When the fronthaul capacity $C$ is sufficient ($C = 10$ bps/Hz), the EE decreases for both IGS and PGS upon increasing $r_{min}$. However, for the limited fronthaul capacity of $C = 5$ bps/Hz, the EE of PGS degrades more dramatically than that of IGS. Additionally, $r_{min} > 0.7$ cannot be attained by PGS. By contrast, IGS attains better fronthaul utilization and thus has higher $r_{min}$.

\begin{figure}[!t]
	\centerline{\includegraphics[width=8cm]{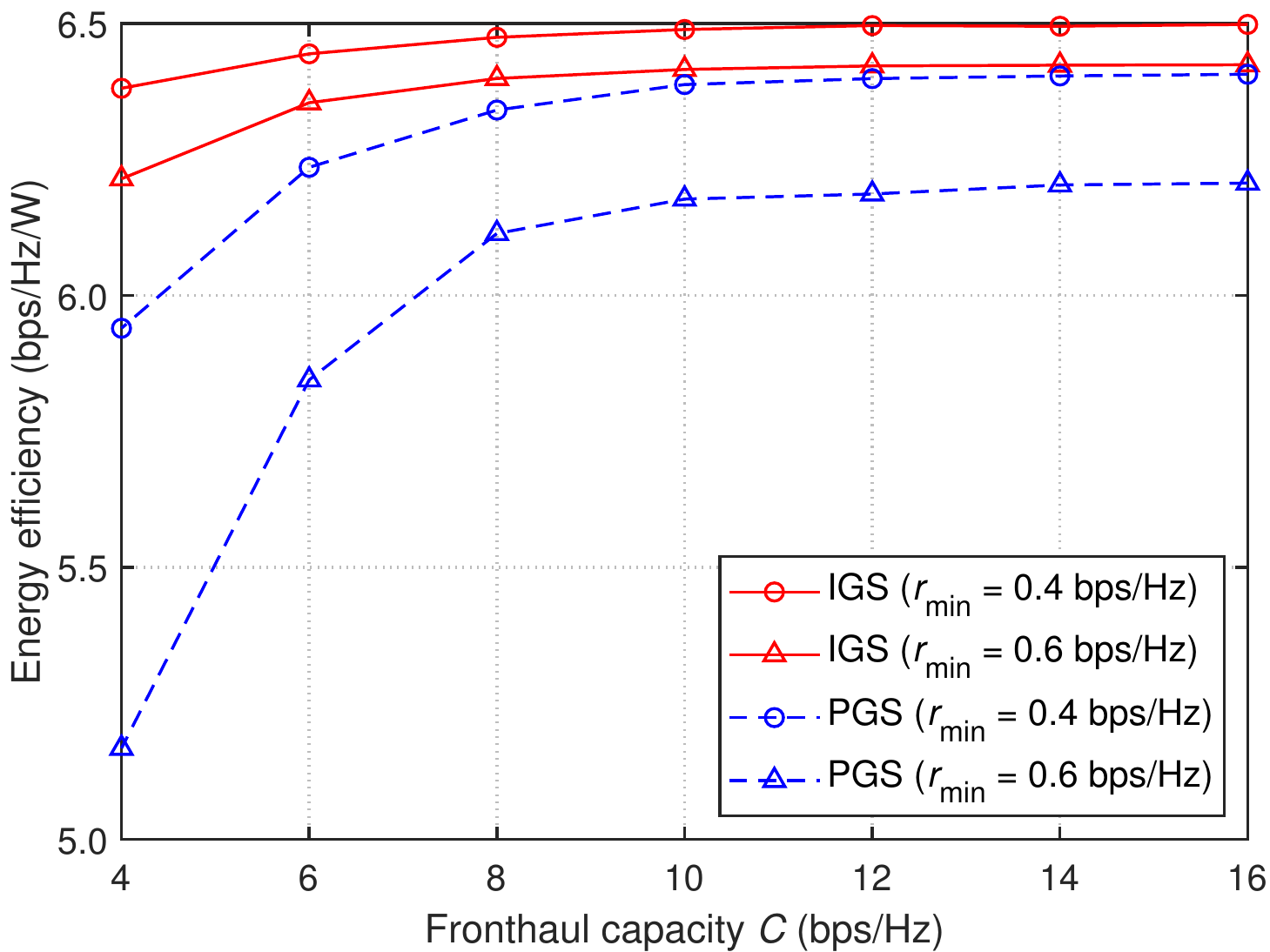}}
	\caption{The energy efficiency versus the fronthaul capacity $C$ using the parameters of Table \ref{table:1}}
	\label{fig:EE_vs_C}
\end{figure}

\begin{figure}[!t]
	\centerline{\includegraphics[width=8cm]{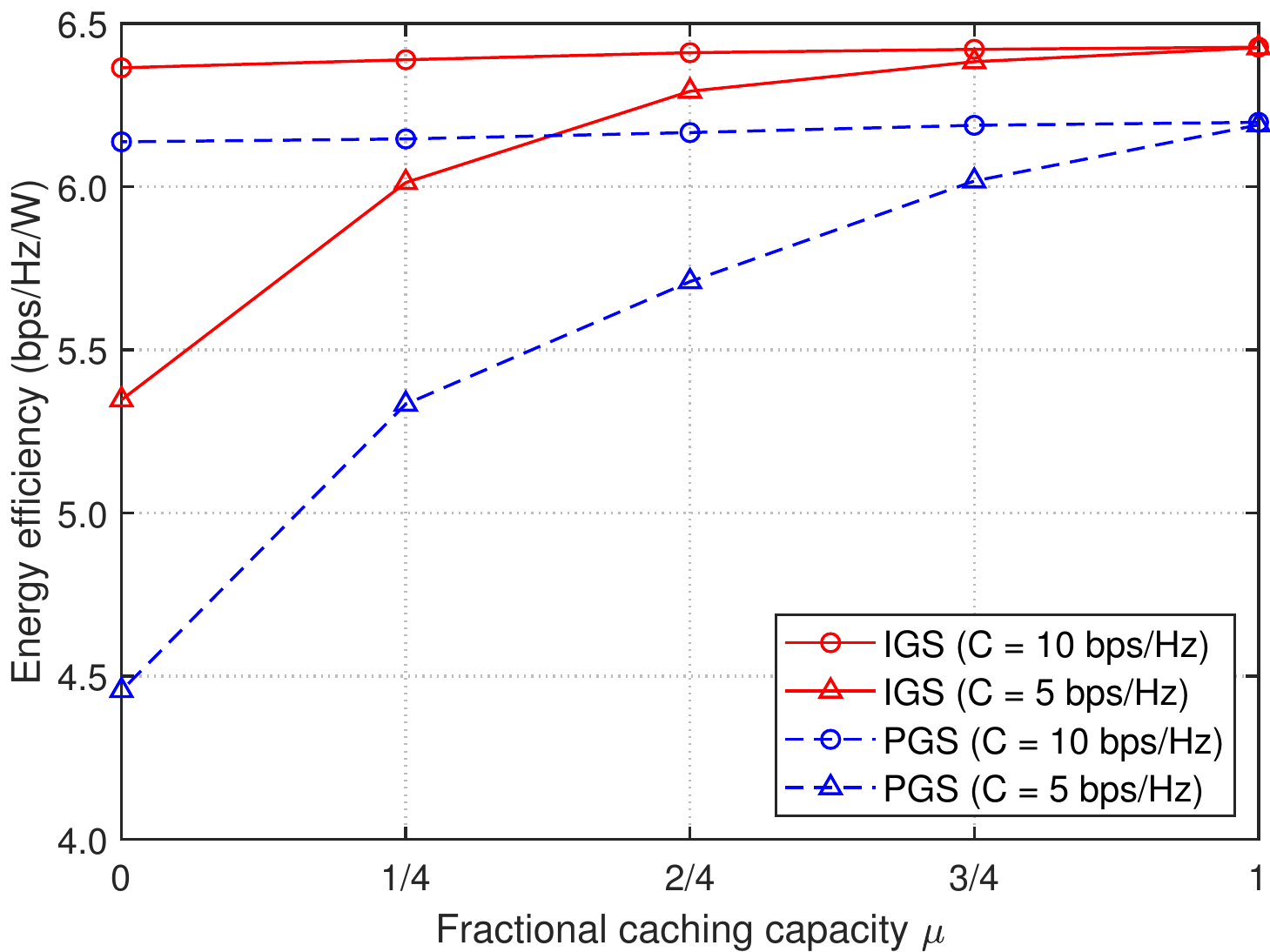}}
	\caption{The energy efficiency versus the fractional caching capacity $\mu$ using the parameters of Table \ref{table:1}}
	\label{fig:EE_vs_mu}
\end{figure}

Fig. \ref{fig:EE_vs_C} plots the achievable EE versus the fronthaul capacity $C$ parameterized by the minimun required rate $r_{min}$ set to $0.4$ bps/Hz and $0.6$ bps/Hz, respectively. The EE exhibits an upward trend upon increasing $C$. In order to reach a higher $r_{min}$, more transmit power is needed. The EE at $r_{min}=0.6$ bps/Hz is lower than that at $r_{min}=0.4$ bps/Hz. Furthermore, IGS outperforms PGS, especially at low fronthaul capacity.

The beneficial effect of the fractional caching capacity $\mu$ on the EE is revealed in Fig. \ref{fig:EE_vs_mu}. We set $r_{min}$ to $0.6$ bps/Hz, and $C$ either to $10$ bps/Hz or to $5$ bps/Hz. The reduction of the fractional caching capacity $\mu$ leads to increasing the number of content downloads via the fronthaul links, inevitably degrading the EE. Observe that the EE of IGS is higher than that of PGS, especially for the lower fronthaul capacity of $C = 5$ bps/Hz. This further illustrates the benefits of IGS. The average number of iterations required for declaring convergence in Fig. \ref{fig:EE_vs_mu} is shown in Table \ref{table:4}.

\begin{figure}[!t]
	\centerline{\includegraphics[width=8cm]{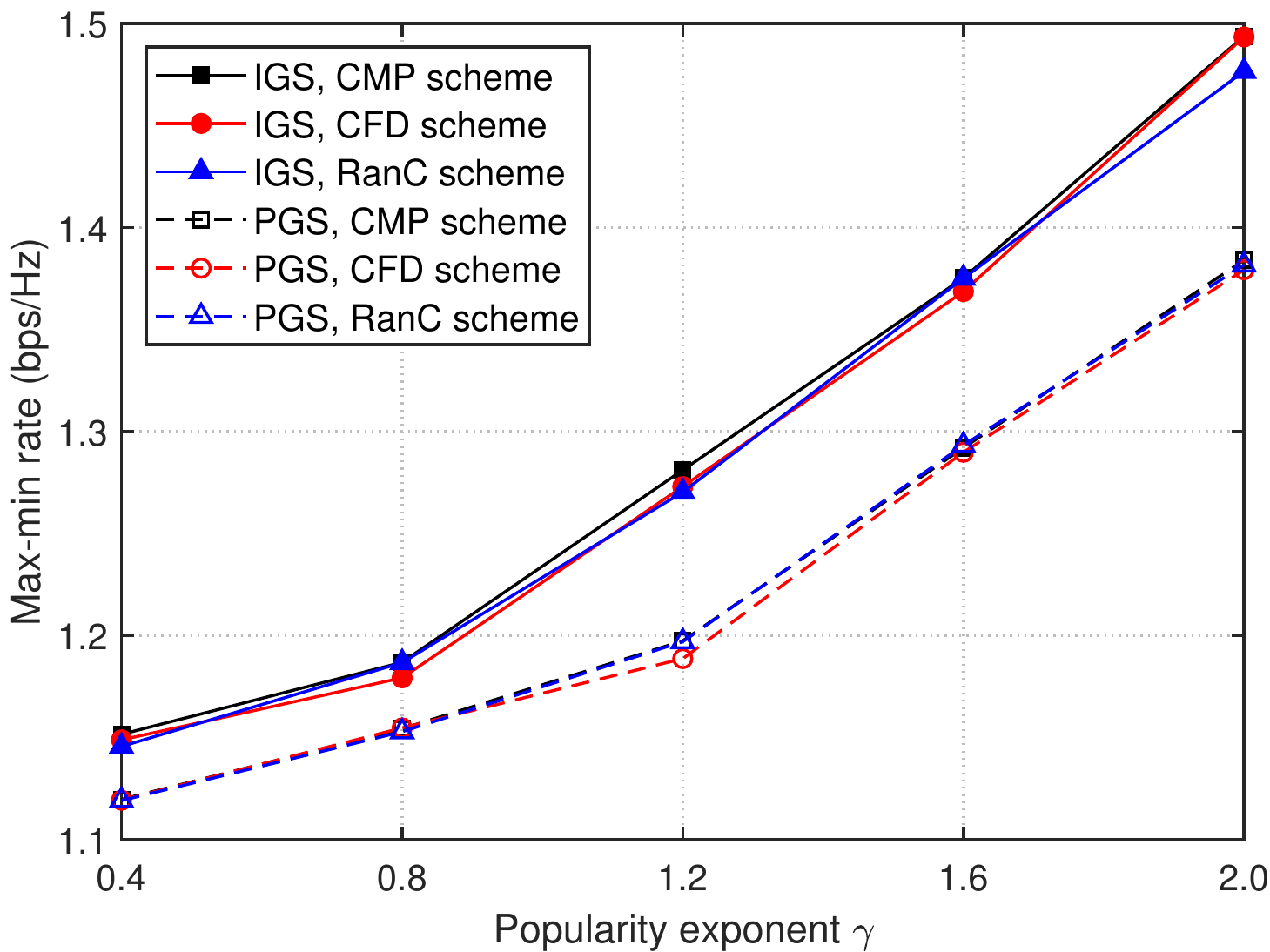}}
	\caption{The max-min rate versus the popularity exponent $\gamma$ without fronthaul constraint using the parameters of Table \ref{table:1}}
	\label{fig:rate_vs_gamma_NoFH}
\end{figure}

\begin{figure}[!t]
	\centerline{\includegraphics[width=8cm]{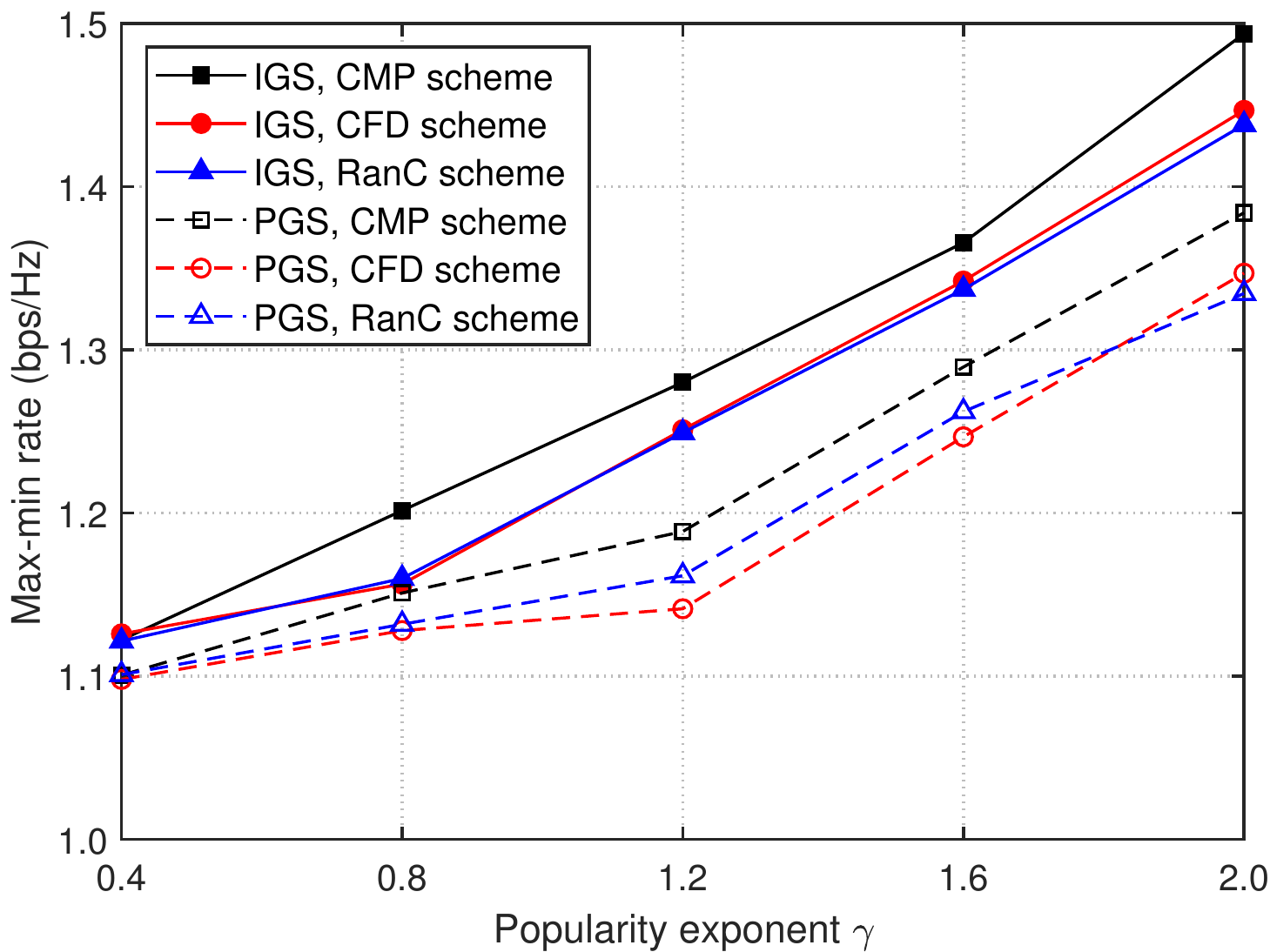}}
	\caption{The max-min rate versus the popularity exponent $\gamma$ with fronthaul constraint using the parameters of Table \ref{table:1}}
	\label{fig:rate_vs_gamma}
\end{figure}

\begin{figure}[!t]
	\centerline{\includegraphics[width=8cm]{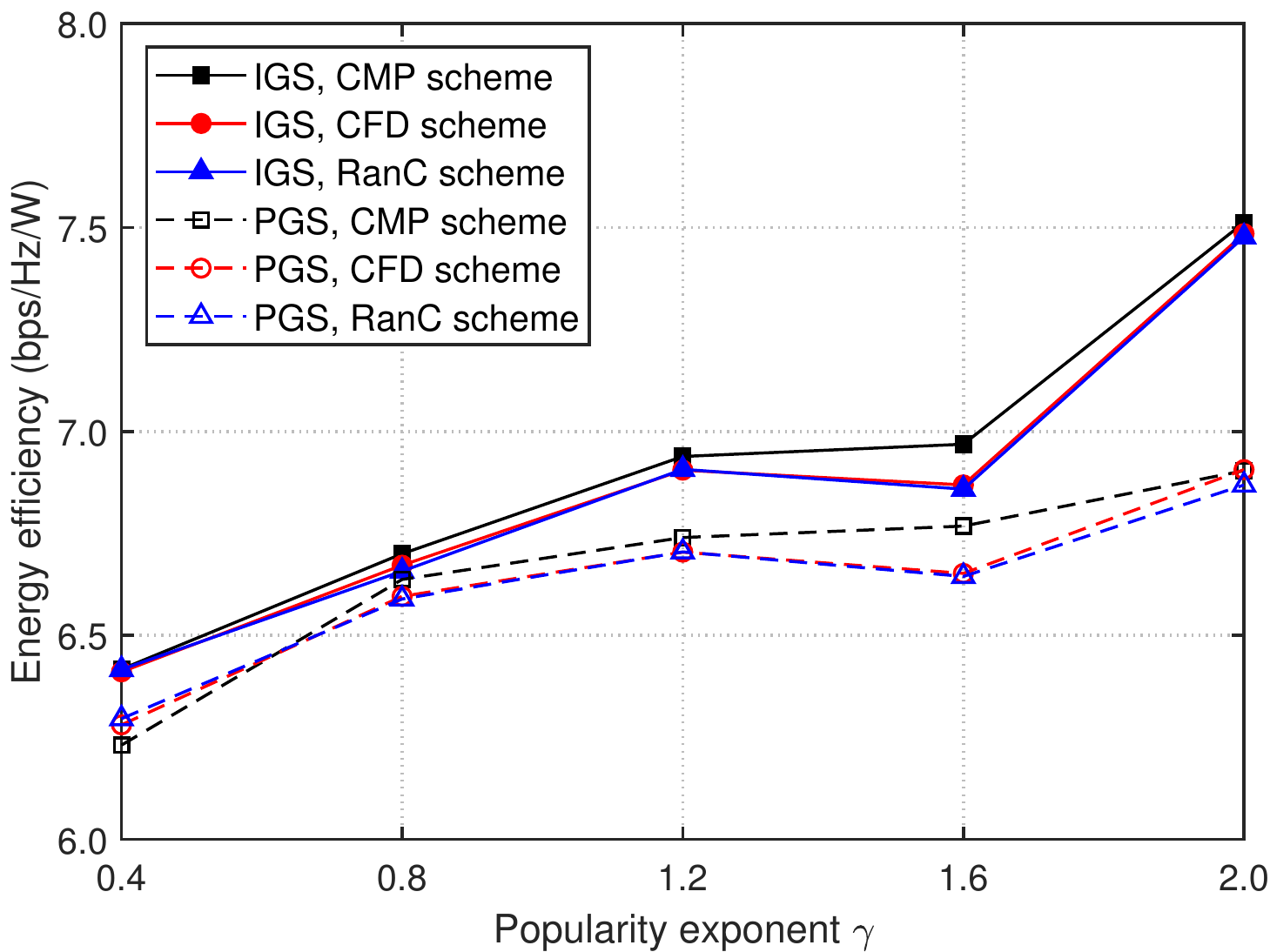}}
	\caption{The energy efficiency versus the popularity exponent $\gamma$ using the parameters of Table \ref{table:1}}
	\label{fig:EE_vs_gamma}
\end{figure}

In Fig. \ref{fig:rate_vs_gamma_NoFH} we characterize the impact of the popularity exponent $\gamma$ on the CMP, CFD, and RanC caching schemes. Fig. \ref{fig:rate_vs_gamma_NoFH} plots the max-min users rate versus the popularity exponent $\gamma$ without any fronthaul constraint. We fix the fractional caching capacity to $\mu = 1/4$. The three caching schemes provide similar performances under the condition that the fronthaul capacity is sufficiently large. The IGS algorithm outperforms the PGS based algorithm for all the three caching schemes. Fig. \ref{fig:rate_vs_gamma} plots the max-min user rate versus the popularity exponent $\gamma$ at the fronthaul constraint of $C = 10$ bps/Hz, where IGS still exhibits better performance. The achievable rate tends to increase with the popularity exponent $\gamma$, because the UE requests are generated by the Zipf distribution, where typically the most popular files are requested at higher $\gamma$, and thus the burden on fronthauling is alleviated.

Fig. \ref{fig:EE_vs_gamma} plots the achievable EE versus the popularity exponent $\gamma$. We fix $r_{min} = 0.6$ bps/Hz, $C = 10$ bps/Hz, and $\mu = 1/4$. As the popularity exponent $\gamma$ is increased, the distribution of requested files is dominated by a small number of the most popular files. Hence less files have to be pre-coded and transmitted over the fronthaul, hence increasing the EE. Since the users' requests are randomly generated, the distribution of the randomly generated requested files becomes unpredictable, especially when $\gamma$ varies from 0.8 to 1.6. The EE of all three caching schemes fluctuates within this range.
As illustrated in Fig. \ref{fig:rate_vs_gamma_NoFH} and Fig. \ref{fig:rate_vs_gamma}, IGS has the highest rate. It also provides high EE as illustrated in Fig. \ref{fig:EE_vs_gamma}. Additionally, the CMP scheme outperforms the other two schemes for high $\gamma$. This is because when the UEs tend to request the most popular files, caching these popular files in the RRHs will put less burden on fronthauling, hence improving the EE. Table \ref{table:5} provides the average number of iterations required for convergence in Fig. \ref{fig:EE_vs_gamma}.

\begin{table}[!t]
	\centering
	\caption{The average number of iterations for implementing IGS and PGS algorithms in obtaining Fig. \ref{fig:EE_vs_mu}}
    \resizebox{1\columnwidth}{!}{
	\begin{tabular}{|l|c|c|c|c|c|}
		\hline
                              & $\mu = 0$ & $\mu = 1/4$ & $\mu = 1/2$ & $\mu = 3/4$ & $\mu = 1$  \\ \hline
        IGS ($C = 10$ bps/Hz) & 10.3        & 14.4      & 12.5      & 13.1      & 13.6               \\
        IGS ($C = 5$ bps/Hz)  & 18.8        & 17.6      & 15.9      & 18.9      & 13.6               \\
        PGS ($C = 10$ bps/Hz) & 8.3         & 8.4       & 8.8       & 9.0       & 8.8                \\
        PGS ($C = 5$ bps/Hz)  & 15.0        & 15.1      & 12.0      & 11.1      & 8.8               \\ \hline
	\end{tabular}}
	\label{table:4}
\end{table}

\begin{table}[!t]
	\centering
	\caption{The average number of iterations for implementing IGS and PGS algorithms in obtaining Fig. \ref{fig:EE_vs_gamma}}
    \resizebox{1\columnwidth}{!}{
	\begin{tabular}{|l|c|c|c|c|c|}
		\hline
                         & $\gamma = 0.4$ & $\gamma = 0.8$ & $\gamma = 1.2$ & $\gamma = 1.6$ & $\gamma = 2.0$ \\ \hline
        IGS, CMP scheme  & 10.6        & 11.1        & 12.0        & 16.1        & 18.9        \\
        IGS, CFD scheme  & 15.0        & 13.9        & 10.9        & 16.3        & 14.9        \\
        IGS, RanC scheme & 9.6         & 10.3        & 11.1        & 16.5        & 14.8        \\
        PGS, CMP scheme  & 9.8         & 9.6         & 14.4        & 16.1        & 14.0        \\
        PGS, CFD scheme  & 8.3         & 8.3         & 10.3        & 15.6        & 13.8        \\
        PGS, RanC scheme & 8.0         & 8.3         & 11.1        & 10.1        & 12.9        \\ \hline
	\end{tabular}}
	\label{table:5}
\end{table}

\section{Conclusions} \label{sec:conclusion}
The performance of content-driven F-RANs is limited by both inter-content
and intra-content interferences.  We mitigated this problem by a new class of RZFB specifically tailored for managing the inter-content interference in the face of limited-capacity backhaul links and minimum user-rate requirements. We also derived expressions having moderate numbers of decision variables in the optimized solutions. The main computational challenges have been resolved by tractable convex quadratic function optimization at a low complexity. Explicitly, we have developed a pair of path-following algorithms for solving our convex quadratic problem by generating an improved feasible point at each iteration. The resultant procedures converged rapidly to locally optimal solutions. An extension enhancing the performance of our data driven multilayered network is under current study.

\section*{Appendix: fundamental inequalities for convex quadratic  approximations}
The following inequality is valid for all positive definite matrices $\bY$ and $\bbY$ of size $n\times n$ \cite{TTN16}:
\begin{align}
&\ln|I_{n}+[\bV]^2\bY^{-1}|\nonumber\\
&\quad\geq \ln|I_{n}+[\bar{V}]^2\bbY^{-1}|-\la[\bar{V}]^2\bbY^{-1}\ra+2\Re\{\la \bar{V}^H\bbY^{-1}\bV\ra\}\nonumber\\
&\qquad-\la \bbY^{-1}-\left([\bar{V}]^2+\bbY\right)^{-1}, [\bV]^2+\bY\ra.\label{fund6}
\end{align}
The particular case of $n=1$ in (\ref{fund6}) is
\begin{align}
\ln\left(1+\frac{|\bv|^2}{\by}\right)&\geq \ln\left(1+\frac{|\bar{v}|^2}{\bar{y}}\right)-\frac{|\bar{v}|^2}{\bar{y}}+2\frac{\Re\{\bar{v}^*\bv\}}{\bar{y}}\nonumber\\
&\quad -\frac{|\bar{v}|^2}{\bar{y}(|\bar{v}|^2+\bar{y})}\left(|\bv|^2+\by\right)\label{fund5}\nonumber\\
&\quad\ \forall \bv\in\mathbb{C}, \by>0 \quad \&\quad \bar{v}\in\mathbb{C}, \bar{y}>0.
\end{align}
Let us define a function
\[
f(\bX,\bY)=\ln|I_m+[\bX]^2([\bY]^2+I_m)^{-1}|
\]
where we have $\bX\in\mathbb{C}^{m\times n}$ and $\bY\in\mathbb{C}^{m\times q}$.
Then for any $\bar{X}\in\mathbb{C}^{m\times n}$ and $\bar{Y}\in\mathbb{C}^{m\times q}$, the following inequality holds
\begin{align}
f(\bX,\bY)&\leq f(\bar{X},\bar{Y})+\la \left(I_m+[\bar{X}]^2+[\bar{Y}]^2\right)^{-1}\ra+\la [\bar{Y}]^2\ra\nonumber\\
&\quad - \la ([\bar{Y}]^2+I_m)^{-1}\ra+\la \left(I_m+[\bar{X}]^2+[\bar{Y}]^2\right)^{-1},\nonumber\\
&\qquad\ [\bX]^2+[\bY]^2\ra-2\Re\{\la \bar{Y}^H\bY\ra\}\nonumber\\
&\quad +\la I_m-([\bar{Y}]^2+I_m)^{-1},[\bY]^2\ra.\label{fund7}
\end{align}
\bibliographystyle{ieeetr}
\bibliography{MCMA_Bib}

\end{document}